\documentclass[11pt]{article}
\usepackage{amsmath,color,graphicx,ifpdf,subfigure,url,wrapfig}
\title{Playing Games with Algorithms: \\ Algorithmic Combinatorial Game Theory%
  \thanks{A preliminary version of this paper appears in
     the \emph{Proceedings of the 26th International Symposium on Mathematical
     Foundations of Computer Science}, Lecture Notes in Computer Science 2136,
     Czech Republic, August 2001, pages 18--32.
     The latest version can be found at
     \url{http://arXiv.org/abs/cs.CC/0106019}.}}
\author{Erik D. Demaine%
  \thanks{MIT Computer Science and Artificial Intelligence Laboratory,
    32 Vassar St., Cambridge, MA 02139, USA, \protect\url{edemaine@mit.edu}}
\and Robert A. Hearn%
  \thanks{Neukom Institute for Computational Sciece, Dartmouth College,
Sudikoff Hall, HB 6255, Hanover, NH 03755, USA,
    \protect\url{robert.a.hearn@dartmouth.edu}}}
\date{}

\advance \topmargin by -\headheight
\advance \topmargin by -\headsep
\evensidemargin \oddsidemargin
\def\captionfont{\rm\small}
\def\captionlabelfont{\bf\small}
{\makeatletter
 \global\let\old@makecaption\@makecaption
 \long\gdef\@makecaption#1#2{%
   \old@makecaption{\captionlabelfont #1}{\captionfont #2}}}

{\makeatletter \gdef\fps@figure{!htbp}}

\let\epsilon=\varepsilon

{\makeatletter
\global\let\GAMEverticalbar=|
\global\let\GAMEstar=*
\catcode`\{=12 \global\let\GAMEleftbrace={
\catcode`\{=13 \catcode`\<=1
\catcode`\}=13 \catcode`\>=2
\catcode`\|=13
\catcode`*=13
\gdef\GAME<\bgroup
  \def\half<\frac12>%
  \def\fuzzy<\,\|\,>%
  \def\up<\uparrow>%
  \catcode`\|=13%
  \def|<\expandafter\@ifnextchar\GAMEverticalbar<\|><\mid>>%
  \catcode`*=13%
  \def*<\GAMEstar\nobreak>%
  \catcode`\{=13%
  \def{<\{>%
  \catcode`\}=13%
  \def}<\}>%
  \GAMEbracket>
\gdef\GAMEbracket[#1]<\ifmmode #1\else $#1$\fi\egroup>
>

\begin{document}
\maketitle

\begin{abstract}
Combinatorial games lead to several interesting, clean problems in algorithms
and complexity theory, many of which remain open.  The purpose of this paper is
to provide an overview of the area to encourage further research.  In
particular, we begin with general background in Combinatorial Game Theory,
which analyzes ideal play in perfect-information games, and Constraint Logic,
which provides a framework for showing hardness.  Then we survey results
about the complexity of determining ideal play in these games, and the related
problems of solving puzzles, in terms of both polynomial-time algorithms and
computational intractability results.  Our review of background and survey of
algorithmic results are by no means complete, but should serve as a useful
primer.
\end{abstract}

\section{Introduction}

Many classic games are known to be computationally intractable (assuming P $\neq$ NP): one-player
puzzles are often NP-complete (as in Minesweeper) or PSPACE-complete (as in Rush Hour),
and two-player games are
often PSPACE-complete (as in Othello) or EXPTIME-complete (as in Checkers,
Chess, and Go).  Surprisingly, many seemingly simple puzzles and games are also
hard.  Other results are positive, proving that some games can
be played optimally in polynomial time.  In some cases, particularly with
one-player puzzles, the computationally tractable games are still interesting
for humans to play.

We begin by reviewing some basics of Combinatorial Game Theory in
Section~\ref{Combinatorial Game Theory}, which gives tools for designing
algorithms, followed by reviewing the relatively new theory of Constraint Logic
in Section~\ref{Constraint Logic}, which gives tools for proving hardness.
In the bulk of this paper,
Sections~\ref{Algorithms for Games}--\ref{Cellular Automata and Life}
survey many of the algorithmic and hardness results for
combinatorial games and puzzles.
Section~\ref{Open Problems} concludes with a small sample of difficult open
problems in algorithmic Combinatorial Game Theory.

Combinatorial Game Theory is to be distinguished from other forms of game
theory arising in the context of economics.  Economic game theory has
many applications in computer science as well, for example, in the context of
auctions \cite{deVries-Vohra-2001} and analyzing behavior on the Internet
\cite{Papadimitriou-2001}.

\section{Combinatorial Game Theory}
\label{Combinatorial Game Theory}

A \emph{combinatorial game} typically involves two players, often called
\emph{Left} and \emph{Right}, alternating play in well-defined \emph{moves}.
However, in the interesting case of a \emph{combinatorial puzzle}, there is
only one player, and for \emph{cellular automata} such as Conway's Game of
Life, there are no players.  In all cases, no randomness or hidden information
is permitted: all players know all information about gameplay (\emph{perfect
information}).  The problem is thus purely strategic: how to best play the game
against an ideal opponent.

It is useful to distinguish several types of two-player perfect-information
games \cite[pp.~14--15]{Berlekamp-Conway-Guy-2001}.  A common assumption is
that the game terminates after a finite number of moves (the game is
\emph{finite} or \emph{short}), and the result is a unique winner.
Of course, there are exceptions: some games (such as Life and Chess)
can be \emph{drawn} out forever, and some games (such as tic-tac-toe and Chess)
define \emph{ties} in certain cases.  However, in the combinatorial-game
setting, it is useful to define the \emph{winner} as the last player who is
able to move; this is called \emph{normal play}.  If, on the other hand,
the winner is the first player who cannot move, this is called
\emph{mis\`ere play}.  (We will normally assume normal play.)
A game is \emph{loopy} if it is possible to return to previously seen positions
(as in Chess, for example).
Finally, a game is called \emph{impartial} if the two players (Left and Right)
are treated identically, that is, each player has the same moves available from
the same game position; otherwise the game is called \emph{partizan}.

A particular two-player perfect-information game without ties or draws can have
one of four \emph{outcomes} as the result of ideal play: player Left wins,
player Right wins, the first player to move wins (whether it is Left or Right),
or the second player to move wins.  One goal in analyzing two-player games is
to determine the outcome as one of these four categories, and to find a
strategy for the winning player to win.  Another goal is to compute
a deeper structure to games described in the remainder of this section,
called the \emph{value} of the game.

A beautiful mathematical theory has been developed for analyzing two-player
combinatorial games.  A new introductory book on the topic is
\emph{Lessons in Play} by Albert, Nowakowski, and Wolfe
\cite{Albert-Nowakowski-Wolfe-2007}; the most comprehensive reference is the
book \emph{Winning Ways} by Berlekamp, Conway, and Guy
\cite{Berlekamp-Conway-Guy-2001}; and a more mathematical presentation is the
book \emph{On Numbers and Games} by Conway \cite{Conway-2001-ONAG}.  See also
\cite{Conway-1977,Fraenkel-1996-overview} for overviews and
\cite{Fraenkel-1994} for a bibliography.  The
basic idea behind the theory is simple: a two-player game can be described
by a rooted tree, where each node has zero or more \emph{left} branches
corresponding to options for player Left to move and zero or more \emph{right}
branches corresponding to options for player Right to move; leaves
correspond to finished games, with the winner determined by either normal
or mis\`ere play.  The interesting parts of Combinatorial Game Theory are the
several methods for manipulating and analyzing such games/trees.  We give a
brief summary of some of these methods in this section.

\subsection{Conway's Surreal Numbers}

A richly structured special class of two-player games are John H. Conway's
\emph{surreal numbers}\footnote{The name ``surreal numbers'' is actually due to
Knuth \cite{Knuth-1974}; see \cite{Conway-2001-ONAG}.}
\cite{Conway-2001-ONAG,Knuth-1974,Gonshor-1986,Alling-1987}, a vast
generalization of the real and ordinal number systems.  Basically, a surreal
number \GAME[{L|R}] is the ``simplest'' number larger than all Left options
(in~$L$) and smaller than all Right options (in~$R$);
for this to constitute a number, all Left and Right options must be numbers,
defining a total order, and each Left option must be less than each Right
option.  See \cite{Conway-2001-ONAG} for more formal definitions.

For example, the simplest number without any larger-than or smaller-than
constraints, denoted \GAME[{|}], is \GAME[0]; the simplest number larger than
\GAME[0] and without smaller-than constraints, denoted \GAME[{0|}], is
\GAME[1]; and the simplest number larger than \GAME[0] and \GAME[1] (or just
\GAME[1]), denoted \GAME[{0,1|}], is \GAME[2].  This method can be used to
generate all natural numbers and indeed all ordinals.  On the other hand, the
simplest number less than \GAME[0], denoted \GAME[{|0}], is \GAME[-1];
similarly, all negative integers can be generated.  Another example is the
simplest number larger than \GAME[0] and smaller than \GAME[1], denoted
\GAME[{0|1}], which is \GAME[\half]; similarly, all dyadic rationals can be
generated.  After a countably infinite number of such construction steps,
all real numbers can be generated; after many more steps, the surreals are all
numbers that can be generated in this way.

Surreal numbers form a field, so in particular they are totally ordered, and
support the operations of addition, subtraction, multiplication, division,
roots, powers, and even integration in many situations.  (For those familiar
with ordinals, contrast with surreals which define $\omega - 1$, $1/\omega$,
$\sqrt \omega$, etc.)
As such, surreal numbers are useful in their own right for cleaner forms of
analysis; see, e.g., \cite{Alling-1987}.

What is interesting about the surreals from the perspective of combinatorial
game theory is that they are a subclass of all two-player perfect-information
games, and some of the surreal structure, such as addition and subtraction,
carries over to general games.  Furthermore, while games are not totally
ordered, they can still be compared to some surreal numbers and,
amazingly, how a game compares to the surreal number \GAME[0] determines
exactly the outcome of the game.  This connection is detailed in the next few
paragraphs.

First we define some algebraic structure of games that carries over from
surreal numbers; see Table~\ref{arithmetic} for formal definitions.
Two-player combinatorial games, or trees, can simply be represented as
\GAME[{L|R}] where, in contrast to surreal numbers, no constraints are placed
on \GAME[L] and \GAME[R].  The \emph{negation} of a game is the result of
reversing the roles of the players Left and Right throughout the game.  The
(\emph{disjunctive}) \emph{sum} of two (sub)games is the game in which, at each
player's turn, the player has a binary choice of which subgame to play, and
makes a move in precisely that subgame.  A partial order is defined on games
recursively: a game $x$ is \emph{less than or equal to} a game $y$ if every
Left option of $x$ is less than $y$ and
every Right option of $y$ is more than $x$.
(Numeric) equality is defined by being both less than or equal to
and more than or equal to.
Strictly inequalities, as used in the definition of less than or equal to,
are defined in the obvious manner.

\begin{table}
\centering
\newenvironment{fboxenv}{\setbox5=\vbox\bgroup\hsize=5.8in}{\egroup\fbox{\box5}}
\begin{fboxenv}%
\begin{minipage}{5.5in}%
  Let \GAME[x = {x^L|x^R}] be a game.
  \def\ITEM{\item[$\bullet$]}
  \begin{itemize}
  \ITEM \GAME[x \leq y] precisely if
        every \GAME[x^L < y] and every \GAME[y^R > x].
  \ITEM \GAME[x = y] precisely if \GAME[x \leq y] and \GAME[x \geq y];
        otherwise \GAME[x \neq y].
  \ITEM \GAME[x < y] precisely if \GAME[x \leq y] and \GAME[x \neq y],
        or equivalently, \GAME[x \leq y] and \GAME[x \not\geq y].
  \ITEM \GAME[-x = {-x^R|-x^L}].
  \ITEM \GAME[x + y = {x^L + y, x + y^L | x^R + y, x + y^R}].
  \ITEM $x$ is \emph{impartial} precisely if $x^L$ and $x^R$ are identical sets
        and recursively every position ($\in x^L = x^R$) is impartial.
  \ITEM A one-pile Nim game is defined by
        \GAME[*n = {*0, \dots, *(n-1) | *0, \dots, *(n-1)}],
        together with \GAME[*0 = 0].
  \end{itemize}
\end{minipage}%
\end{fboxenv}%
\vspace{2ex} 
\caption{\label{arithmetic}
  Formal definitions of some algebra on two-player perfect-information games.
  In particular, all of these notions apply to surreal numbers.}
\end{table}

Note that while \GAME[{-1|1} = 0 = {|}] in terms of numbers, \GAME[{-1|1}] and
\GAME[{|}] denote different games (lasting 1 move and 0 moves, respectively),
and in this sense are \emph{equal} in \emph{value} but not \emph{identical}
symbolically or game-theoretically.  Nonetheless, the games
\GAME[{-1|1}] and \GAME[{|}] have the same outcome: the second player to move
wins.

Amazingly, this holds in general: two equal numbers represent games with equal
outcome (under ideal play).  In particular, all games equal to 0 have the
outcome that the second player to move wins.  Furthermore, all games equal to a
positive number have the outcome that the Left player wins; more generally, all
positive games (games larger than \GAME[0]) have this outcome.
Symmetrically, all negative games have the outcome that the Right player wins
(this follows automatically by the negation operation).
Examples of zero, positive, and negative games are the surreal numbers
themselves; an additional example is described below.

There is one outcome not captured by the characterization into zero, positive,
and negative games: the first player to move wins.
To find such a game we must
obviously look beyond the surreal numbers.  Furthermore, we must look for games
$G$ that are incomparable with zero (none of $G = 0$, $G < 0$, or $G > 0$
hold); such games are called \emph{fuzzy} with $0$, denoted \GAME[G \fuzzy 0].

An example of a game that is not a surreal number is \GAME[{1|0}]; there fails
to be a number strictly between $1$ and $0$ because $1 \geq 0$.
Nonetheless, \GAME[{1|0}] is a game: Left has a single move leading
to game \GAME[1], from which Right cannot move, and Right has a single move
leading to game \GAME[0], from which Left cannot move.  Thus, in either case,
the first player to move wins.  The claim above implies that \GAME[{1|0}
\fuzzy 0].
Indeed, \GAME[{1|0} \fuzzy x] for all surreal numbers $x$,
$0 \leq x \leq 1$.  In contrast, \GAME[x < {1|0}] for all $x < 0$
and \GAME[{1|0} < x] for all $1 < x$.  In general it holds that a game is fuzzy
with some surreal numbers in an interval $[-n,n]$ but comparable with all
surreals outside that interval.
Another example of a game that is not a number is \GAME[{2|1}], which is
positive ($>0$), and hence Right wins, but fuzzy with numbers in the range
$[1,2]$.

For brevity we omit many other useful notions in Combinatorial Game Theory,
such as additional definitions of summation,
super-infinitesimal games \GAME[*] and \GAME[\up],
mass, temperature, thermographs,
the simplest form of a game, remoteness, and suspense;
see \cite{Berlekamp-Conway-Guy-2001,Conway-2001-ONAG}.

\subsection{Sprague-Grundy Theory}

A celebrated result in Combinatorial Game Theory is the characterization of
impartial two-player perfect-information games, discovered independently in the
1930's by Sprague \cite{Sprague-1935} and Grundy \cite{Grundy-1939}.
Recall that a game is \emph{impartial} if it does not
distinguish between the players Left and
Right (see Table~\ref{arithmetic} for a more formal definition).
The Sprague-Grundy theory
\cite{Sprague-1935,Grundy-1939,Conway-2001-ONAG,Berlekamp-Conway-Guy-2001}
states that every finite impartial game is equivalent to an instance of the
game of Nim, characterized by a single natural number~$n$.
This theory has since been generalized to all impartial games by generalizing
Nim to all ordinals~$n$; see \cite{Conway-2001-ONAG,Smith-1966}.

\emph{Nim} \cite{Bouton-1901} is a game played with several
\emph{heaps}, each with a certain number of tokens.  A Nim game with a single
heap of size $n$ is denoted by \GAME[*n] and is called a \emph{nimber}.  During
each move a player can pick any pile and reduce it to any smaller nonnegative
integer size.  The game ends when all piles have size $0$.  Thus, a single pile
\GAME[*n] can be reduced to any of the smaller piles \GAME[*0], \GAME[*1],
\dots, \GAME[*(n-1)].  Multiple piles in a game of Nim are independent, and
hence any game of Nim is a sum of single-pile games \GAME[*n] for various
values of $n$.  In fact, a game of Nim with $k$ piles of sizes $n_1$, $n_2$,
\dots, $n_k$ is equivalent to a one-pile Nim game \GAME[*n], where $n$ is the
binary XOR of $n_1$, $n_2$, \dots, $n_k$.  As a consequence, Nim can be played
optimally in polynomial time (polynomial in the encoding size of the pile
sizes).

Even more surprising is that \emph{every} impartial two-player
perfect-information game has the same value as a single-pile Nim game,
\GAME[*n] for some $n$.  The number $n$ is called the \emph{G-value},
\emph{Grundy-value}, or \emph{Sprague-Grundy function} of the game.  It is easy
to define: suppose that game $x$ has $k$ options $y_1, \dots, y_k$ for the
first move (independent of which player goes first).  By induction, we can
compute $y_1 = *n_1$, \dots, $y_k = *n_k$.  The theorem is that $x$ equals $*n$
where $n$ is the smallest natural number not in the set $\{n_1, \dots, n_k\}$.
This number $n$ is called the \emph{minimum excluded value} or \emph{mex} of
the set.  This description has also assumed that the game is finite, but this
is easy to generalize \cite{Conway-2001-ONAG,Smith-1966}.

The Sprague-Grundy function can increase by at most $1$ at each level of the
game tree, and hence the resulting nimber is linear in the maximum number of
moves that can be made in the game; the encoding size of the nimber is only
logarithmic in this count.  Unfortunately, computing the Sprague-Grundy
function for a general game by the obvious method uses time linear in the
number of possible states, which can be exponential in the nimber itself.

Nonetheless, the Sprague-Grundy theory is extremely helpful for analyzing
impartial two-player games, and for many games there is an efficient algorithm
to determine the nimber.  Examples include Nim itself, Kayles, and various
generalizations \cite{Guy-Smith-1956}; and
Cutcake and Maundy Cake \cite[pp.~24--27]{Berlekamp-Conway-Guy-2001}.
In all of these examples, the Sprague-Grundy function has a succinct
characterization (if somewhat difficult to prove); it can also be easily
computed using dynamic programming.

The Sprague-Grundy theory seems difficult to generalize to the superficially
similar case of mis\`ere play, where the goal is to be the first player
unable to move.  Certain games have been solved in this context over the years,
including Nim \cite{Bouton-1901};
see, e.g., \cite{Ferguson-1974,Grundy-Smith-1956}.
Recently a general theory has emerged for tackling mis\`ere combinatorial
games, based on commutative monoids called ``mis\`ere quotients''
that localize the problem to certain restricted game scenarios.
This theory was introduced by Plambeck \cite{Plambeck-2005}
and further developed by Plambeck and Siegel \cite{Plambeck-Siegel-2007}.
For good descriptions of the theory, see Plambeck's survey
\cite{Plambeck-2008-survey}, Siegel's lecture notes \cite{Siegel-2006-misere},
and a webpage devoted to the topic \cite{miseregames.org}.

\subsection{Strategy Stealing}
\label{Strategy Stealing}

Another useful technique in Combinatorial Game Theory for proving that a
particular player must win is \emph{strategy stealing}.  The basic idea is to
assume that one player has a winning strategy, and prove that in fact the other
player has a winning strategy based on that strategy.  This contradiction
proves that the second player must in fact have a winning strategy.  An example
of such an argument is given in Section~\ref{Hex}.
Unfortunately, such a proof by contradiction gives no indication of what the
winning strategy actually is, only that it exists.  In many situations, such as
the one in Section~\ref{Hex}, the winner is known but no polynomial-time
winning strategy is known.

\subsection{Puzzles}
\label{Puzzles}

There is little theory for analyzing combinatorial puzzles (one-player games)
along the lines of the two-player theory summarized in this section.
We present one such viewpoint here.
In most puzzles, solutions subdivide into a sequence of moves.
Thus, a puzzle can be viewed as a tree, similar to a two-player game
except that edges are not distinguished between Left and Right.
With the view that the game ends only when the puzzle is solved,
the goal is then to reach a position from which there are no valid moves
(normal play).  
Loopy puzzles are common; to be more explicit, repeated subtrees can be
converted into self-references to form a directed graph,
and losing terminal positions can be given explicit loops to themselves.

A consequence of the above view is that a puzzle is basically an impartial
two-player game except that we are not interested in the outcome from two
players alternating in moves.  Rather, questions of interest in the context of
puzzles are (a)~whether a given puzzle is solvable, and (b)~finding the
solution with the fewest moves.  An important open direction of research is to
develop a general theory for resolving such questions, similar to the
two-player theory. 

\section{Constraint Logic}
\label{Constraint Logic}

Combinatorial Game Theory provides a theoretical framework for giving positive
algorithmic results for games, but does not naturally accommodate puzzles.
In contrast, negative algorithmic results---hardness and completeness within
computational complexity classes---are more uniform: puzzles and games
have analogous prototypical proof structures.
Furthermore, a relatively new theory called Constraint Logic attempts
to tie together a wide range of hardness proofs for both puzzles and games.

Proving that a problem is hard within a particular complexity class
(like NP, PSPACE, or EXPTIME) almost always involves a reduction to the problem
from a known hard problem within the class.
For example, the canonical problem to reduce from for NP-hardness is
Boolean Satisfiability (SAT) \cite{Cook-1971}.
Reducing SAT to a puzzle of interest proves that that puzzle is NP-hard.
Similarly, the canonical problem to reduce from for PSPACE-hardness is
Quantified Boolean Formulas (QBF) \cite{Stockmeyer-Meyer-1973}.

\emph{Constraint Logic} \cite{Demaine-Hearn-2008} is a useful tool for
showing hardness of games and puzzles in a variety of settings
that has emerged in recent years.
Indeed, many of the hardness results mentioned in this survey are based on
reductions from Constraint Logic.
Constraint Logic is a family of games where players reverse edges on a
planar directed graph while satisfying vertex in-flow constraints.
Each edge has a weight of $1$ or~$2$.  Each vertex has degree $3$ and requires
that the sum of the weights of inward-directed edges is at least~$2$.
Vertices may be restricted to two types:
\emph{\textsc{And}} vertices have incident edge weights of $1$, $1$, and~$2$;
and
\emph{\textsc{Or}} vertices have incident edge weights of $2$, $2$, and~$2$.
A player's goal is to eventually reverse a given edge.

This game family can be interpreted in many game-theoretic settings,
ranging from zero-player automata to multiplayer games with hidden information.
In particular, there are natural versions of Constraint Logic corresponding to
one-player games (puzzles) and two-player games, both of bounded and unbounded
length.  (Here we refer to whether the length of the game is bounded by a
polynomial function of the board size.
Typically, bounded games are nonloopy while unbounded games are loopy.)
These games have the expected complexities: one-player bounded games are
NP-complete; one-player unbounded games and two-player bounded games are
PSPACE-complete; and two-player unbounded games are EXPTIME-complete.

What makes Constraint Logic specially suited for game and puzzle reductions is
that the problems are already in form similar to many games.
In particular, the fact that the games are played on planar graphs means that
the reduction does not usually need a crossover gadget,
whereas historically crossover gadgets have often been the complex crux of
a game hardness proof.

Historically, Constraint Logic arose as a simplification of the ``Generalized
Rush-Hour Logic'' of Flake and Baum \cite{Flake-Baum-2002}.
The resulting one-player unbounded setting, called
\emph{Nondeterministic Constraint Logic}
\cite{Hearn-Demaine-2002, Hearn-Demaine-2005},
was later generalized to other game categories
\cite{Hearn-2006, Demaine-Hearn-2008}.

\section{Algorithms for Two-Player Games}
\label{Algorithms for Games}

Many bounded-length two-player games are PSPACE-complete.  This is fairly natural
because games are closely related to Boolean expressions with alternating
quantifiers (for which deciding satisfiability is PSPACE-complete): there
exists a move for Left such that, for all moves for Right, there exists another
move for Left, etc.  A PSPACE-completeness result has two consequences.  First,
being in PSPACE means that the game can be played optimally, and typically all
positions can be enumerated, using possibly exponential time but only
polynomial space.  Thus such games lend themselves to a somewhat reasonable
exhaustive search for small enough sizes.
Second, the games cannot be solved
in polynomial time unless P = PSPACE, which is even ``less likely'' than P
equaling~NP.

On the other hand, unbounded-length two-players games are often EXPTIME-complete.  Such a
result is one of the few types of true lower bounds in complexity theory,
implying that all algorithms require exponential time in the worst case.

In this section we briefly survey many of these complexity results and related
positive results.
See also \cite{Eppstein-cgt-hard} for a related survey
and \cite{Fraenkel-1994} for a bibliography.

\begin{wrapfigure}{r}{2in}
  \centering
  \vspace*{-\intextsep}
  \vspace*{-4ex}
  \includegraphics[width=2in]{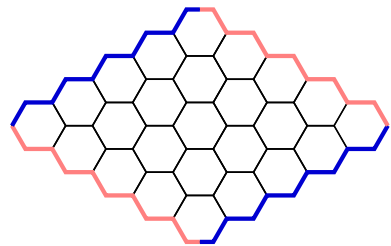}
  \caption{A $5 \times 5$ Hex board.}
  \vspace*{-2ex}
  \label{hex5}
\end{wrapfigure}

\subsection{Hex}
\label{Hex}

Hex \cite[pp.~743--744]{Berlekamp-Conway-Guy-2001} is a game designed by Piet
Hein and played on a diamond-shaped hexagonal board; see Figure~\ref{hex5}.
Players take turns filling in empty hexagons with their color.  The goal of a
player is to connect the opposite sides of their color with hexagons of their
color.  (In the figure, one player is solid and the other player is dotted.)  A
game of Hex can never tie, because if all hexagons are colored arbitrarily,
there is precisely one connecting path of an appropriate color between opposite
sides of the board.

John Nash \cite[p.~744]{Berlekamp-Conway-Guy-2001}
proved that the first player to
move can win by using a strategy-stealing argument (see Section~\ref{Strategy
Stealing}).
Suppose that the second player has a winning strategy, and assume
by symmetry that Left goes first.  Left selects the first hexagon arbitrarily.
Now Right is to move first and Left is effectively the second player.  Thus,
Left can follow the winning strategy for the second player, except that Left
has one additional hexagon.  But this additional hexagon can only help Left: it
only restricts Right's moves, and if Left's strategy suggests filling the
additional hexagon, Left can instead move anywhere else.  Thus, Left has a
winning strategy, contradicting that Right did, and hence the first player has
a winning strategy.
However, it remains open to give a polynomial characterization of a
winning strategy for the first player.

In perhaps the first PSPACE-hardness result for ``interesting'' games, Even and
Tarjan \cite{Even-Tarjan-1976} proved that a generalization of Hex to graphs is
PSPACE-complete, even for maximum-degree-$5$ graphs.  Specifically, in this
graph game, two vertices are initially colored Left, and players take turns
coloring uncolored vertices in their own color.  Left's goal is to connect the
two initially Left vertices by a path, and Right's goal is to prevent such a
path.  Surprisingly, the closely related problem in which players color
\emph{edges} instead of vertices can be solved in polynomial time; this game is
known as the \emph{Shannon switching game} \cite{Bruno-Weinberg-1970}.  A
special case of this game is \emph{Bridgit} or \emph{Gale}, invented by David
Gale \cite[p.~744]{Berlekamp-Conway-Guy-2001}, in which the graph is a square
grid and Left's goal is to connect a vertex on the top row with a vertex on the
bottom row.  However, if the graph in Shannon's switching game has directed
edges, the game again becomes PSPACE-complete \cite{Even-Tarjan-1976}.

A few years later, Reisch \cite{Reisch-1981} proved the stronger result that
determining the outcome of a position in Hex is PSPACE-complete on a normal
diamond-shaped board.  The proof is quite different from the general graph
reduction of Even and Tarjan \cite{Even-Tarjan-1976}, but the main milestone is
to prove that Hex is PSPACE-complete for planar graphs.

\subsection{More Games on Graphs:
            Kayles, Snort, Geography, Peek, and Interactive Hamiltonicity}
\label{More Games on Graphs}

The second paper to prove PSPACE-hardness of ``interesting'' games is by
Schaefer \cite{Schaefer-1978-games}.  This work proposes over a dozen games and
proves them PSPACE-complete.  Some of the games involve propositional formulas,
others involve collections of sets, but perhaps the most interesting are those
involving graphs.  Two of these games are generalizations of ``Kayles'', and
another is a graph-traversal game called Edge Geography.

\emph{Kayles} \cite[pp.~81--82]{Berlekamp-Conway-Guy-2001} is an impartial
game, designed independently by Dudeney and Sam Loyd, in which bowling pins are
lined up on a line.  Players take turns \emph{bowling} with the property that
exactly one or exactly two adjacent pins are knocked down (removed) in each
move.  Thus, most moves split the game into a sum of two subgames.  Under
normal play, Kayles can be solved in polynomial time using the Sprague-Grundy
theory; see \cite[pp.~90--91]{Berlekamp-Conway-Guy-2001},
\cite{Guy-Smith-1956}.

\emph{Node Kayles} is a generalization of Kayles to graphs in which each bowl
``knocks down'' (removes) a desired vertex and all its neighboring vertices.
(Alternatively, this game can be viewed as two players finding an independent
set.)  Schaefer \cite{Schaefer-1978-games} proved that deciding the outcome of
this game is PSPACE-complete.  The same result holds for a partizan version of
node Kayles, in which every node is colored either Left or Right and only the
corresponding player can choose a particular node as the primary target.

\emph{Geography} is another graph game, or rather game family, 
that is special from a techniques point of view: it
has been used as the basis of many other PSPACE-hardness reductions for games
described in this section.  The motivating example of the game is
players taking turns naming distinct geographic locations, each starting with the same
letter with which the previous name ended.  More generally, Geography consists
of a directed graph with one node initially containing a token.  Players
take turns moving the token along a directed edge. In \emph{Edge Geography},
that edge is then erased; in \emph{Vertex Geography}, the vertex moved from is
then erased. (Confusingly, in the literature, each of these variants is frequently
referred to as simply ``Geography'' or ``Generalized Geography''.)

Schaefer \cite{Schaefer-1978-games} established that Edge Geography
(a game  suggested by R. M. Karp) is
PSPACE-complete;  Lichtenstein and Sipser \cite{Lichtenstein-Sipser-1980} showed that Vertex Geography
(which more closely matches the motivating example above) is also 
PSPACE-complete. Nowakowski and Poole \cite{Nowakowski-Poole-1996} have solved
special cases of Vertex Geography when the graph is a product of two cycles.

One may also consider playing either Geography game on an undirected graph. 
Fraenkel, Scheinerman, and Ullman \cite{Fraenkel-Scheinerman-Ullman-1993}
show that Undirected Vertex Geography can be solved in polynomial time, whereas
Undirected Edge Geography is PSPACE-complete, even for planar graphs with maximum
degree 3. If the graph is bipartite then 
Undirected Edge Geography is also solvable in polynomial time.

One consequence of partizan node Kayles being PSPACE-hard is that deciding the
outcome in Snort is PSPACE-complete on general graphs
\cite{Schaefer-1978-games}.  \emph{Snort}
\cite[pp.~145--147]{Berlekamp-Conway-Guy-2001} is a game designed by S. Norton
and normally played on planar graphs (or planar maps).  In any case, players
take turns coloring vertices (or faces) in their own color such that only equal
colors are adjacent.

Generalized hex (the vertex Shannon switching game), node Kayles, and Vertex Geography
have also been analyzed recently in the context of parameterized complexity.
Specifically, the problem of deciding whether the first player can win within
$k$ moves, where $k$ is a parameter to the problem, is AW[$*$]-complete
\cite[ch.~14]{Downey-Fellows-1997}.


Stockmeyer and Chandra \cite{Stockmeyer-Chandra-1979} were the first to prove
combinatorial games to be EXPTIME-hard.  They established EXPTIME-completeness
for a class of logic games and two graph games.  Here we describe an example
of a logic game in the class, and one of the graph games; the other graph game
is described in the next section.
One logic game, called Peek, involves a box containing several parallel
rectangular plates.
Each plate (1)~is colored either Left or Right except for one ownerless plate,
(2)~has circular holes carved in particular (known) positions, and (3)~can be
slid to one of two positions (fully in the box or partially outside the box).
Players take turns either passing or changing the position of one of their
plates.  The winner is the first player to cause a hole in every plate to be
aligned along a common vertical line.  A second game involves a graph in which
some edges are colored Left and some edges are colored Right, and initially
some edges are ``in'' while the others are ``out''.  Players take turns either
passing or changing one edge from ``out'' to ``in'' or vice versa.  The winner
is the first player to cause the graph of ``in'' edges to have a Hamiltonian
cycle.
(Both of these games can be rephrased under normal play by defining there to be
no valid moves from positions having aligned holes or Hamiltonian cycles.)

\subsection{Games of Pursuit: Annihilation, Remove, Capture, Contrajunctive, Blocking, Target, and Cops and Robbers}

The next suite of graph games essentially began study in 1976 when Fraenkel and
Yesha \cite{Fraenkel-Yesha-1976} announced that a certain impartial
annihilation game could be played optimally in polynomial time.  Details
appeared later in \cite{Fraenkel-Yesha-1982}; see also \cite{Fraenkel-1974}.
The game was proposed by John Conway and is played on an
arbitrary directed graph in which some of the vertices contain a token.
Players take turns selecting a token and moving it along an edge; if this
causes the token to occupy a vertex already containing a token, both tokens are
\emph{annihilated} (removed).  The winner is determined by normal play if all
tokens are annihilated, except that play may be drawn out indefinitely.
Fraenkel and Yesha's result \cite{Fraenkel-Yesha-1982} is that the outcome of
the game can be determined and (in the case of a winner) a winning strategy of
$O(n^5)$ moves can be computed in $O(n^6)$ time, where $n$ is the number of
vertices in the graph.

A generalization of this impartial game, called \emph{Annihilation}, is when
two (or more) types of tokens are distinguished, and each type of token can
travel along only a certain subset of the edges.  As before, if a token is
moved to a vertex containing a token (of any type), both tokens are
annihilated.  Determining the outcome of this game was proved NP-hard
\cite{Fraenkel-Yesha-1979} and later PSPACE-hard
\cite{Fraenkel-Goldschmidt-1987}.  For acyclic graphs, the problem is
PSPACE-complete \cite{Fraenkel-Goldschmidt-1987}.  The precise complexity for
cyclic graphs remains open.
Annihilation has also been studied under mis\`ere play \cite{Ferguson-1984}.

A related impartial game, called \emph{Remove}, has the same rules as
Annihilation except that when a token is moved to a vertex containing another
token, only the moved token is removed.  This game was also proved NP-hard
using a reduction similar to that for Annihilation \cite{Fraenkel-Yesha-1979},
but otherwise its complexity seems open.  The analogous impartial game
in which just the \emph{unmoved} token is removed, called \emph{Hit}, is
PSPACE-complete for acyclic graphs \cite{Fraenkel-Goldschmidt-1987},
but its precise complexity remains open for cyclic graphs.

A partizan version of Annihilation is \emph{Capture}, in which the two types of
tokens are assigned to corresponding players.  Left can only move a Left token,
and only to a position that does not contain a Left token.  If the position
contains a Right token, that Right token is \emph{captured} (removed).
Unlike Annihilation, Capture allows all tokens to travel along all edges.
Determining the outcome of Capture was proved NP-hard
\cite{Fraenkel-Yesha-1979} and later EXPTIME-complete
\cite{Goldstein-Reingold-1995}.  For acyclic graphs the game is
PSPACE-complete \cite{Goldstein-Reingold-1995}.

A different partizan version of Annihilation is \emph{Contrajunctive}, in which
players can move both types of tokens, but each player can use only a certain
subset of the edges.  This game is NP-hard even for acyclic graphs
\cite{Fraenkel-Yesha-1979} but otherwise its complexity seems open.

The \emph{Blocking} variations of Annihilation disallow a token to be moved to
a vertex containing another token.  Both variations are partizan and played
with tokens on directed graph.  In \emph{Node Blocking}, each token is assigned
to one of the two players, and all tokens can travel along all edges.
Determining the outcome of this game was proved NP-hard
\cite{Fraenkel-Yesha-1979}, then PSPACE-hard \cite{Fraenkel-Goldschmidt-1987},
and finally EXPTIME-complete \cite{Goldstein-Reingold-1995}.  Its status for
acyclic graphs remains open.  In \emph{Edge Blocking}, there is only one type
of token, but each player can use only a subset of the edges.  Determining the
outcome of this game is PSPACE-complete for acyclic
graphs \cite{Fraenkel-Goldschmidt-1987}.  Its precise complexity for general
graphs remains open.

A generalization of Node Blocking is \emph{Target}, in which some nodes are
marked as \emph{targets} for each player, and players can additionally win by
moving one of their tokens to a vertex that is one of their targets.  When no
nodes are marked as targets, the game is the same as Blocking and hence
EXPTIME-complete by \cite{Goldstein-Reingold-1995}.  In fact, general Target
was proved EXPTIME-complete earlier by Stockmeyer and Chandra
\cite{Stockmeyer-Chandra-1979}.  Surprisingly, even the special case in which
the graph is acyclic and bipartite and only one player has targets
is PSPACE-complete \cite{Goldstein-Reingold-1995}.  (NP-hardness of this case
was established earlier \cite{Fraenkel-Yesha-1979}.)

A variation on Target is \emph{Semi-Partizan Target}, in which both players can
move all tokens, yet Left wins if a Left token reaches a Left target,
independent of who moved the token there.  In addition, if a token is moved to
a nontarget vertex containing another token, the two tokens are annihilated.
This game is EXPTIME-complete \cite{Goldstein-Reingold-1995}.
While this game may seem less natural than the others, it was intended as a
step towards the resolution of Annihilation.

Many of the results described above from \cite{Goldstein-Reingold-1995} are
based on analysis of a more complex game called \emph{Pursuit} or \emph{Cops
and Robbers}.  One player, the \emph{robber}, has a single token;
and the other player, the \emph{cops}, have $k$ tokens.
Players take turns moving all of their tokens along edges in a directed graph.
The cops win if at the end of any move the robber occupies the same vertex
as a cop, and the robber wins if play can be forced to draw out forever.
In the case of a single cop ($k=1$), there is a simple polynomial-time
algorithm, and in general, many versions of the game are EXPTIME-complete;
see \cite{Goldstein-Reingold-1995} for a summary.
For example, EXPTIME-completeness holds even for undirected graphs,
and for directed graphs in which cops and robbers can choose their initial
positions.  For acyclic graphs, Pursuit is PSPACE-complete
\cite{Goldstein-Reingold-1995}.

\begin{wrapfigure}{r}{1.8in}
  \centering
  \vspace*{-\intextsep}
  \vspace{-7ex}
  \includegraphics[width=1.8in]{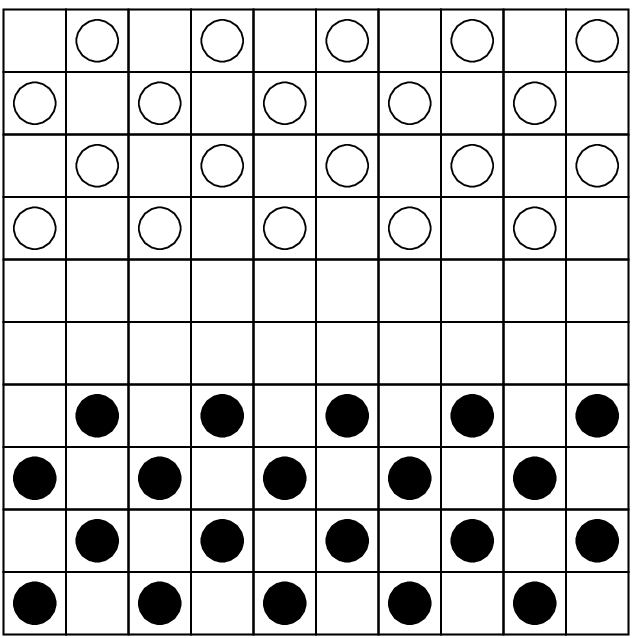}
  \caption{A natural starting configuration for $10 \times 10$ Checkers,
    from \protect\cite{Fraenkel-Garey-Johnson-Schaefer-Yesha-1978}.}
  \vspace{-3ex}
  \label{10x10 checkers starting}
\end{wrapfigure}

\subsection{Checkers (Draughts)}

The standard $8 \times 8$ game of Checkers (Draughts), like many classic games,
is finite and hence can be played optimally in constant time (in theory).
Indeed, Schaeffer et al.~\cite{Schaeffer-Burch-Bjoernsson-Kishimoto-Mueller-Lake-Lu-Sutphen-2007}
recently computed that optimal play leads to a draw from the initial
configuration (other configurations remain unanalyzed).
The outcome of playing in a general $n \times n$ board from a natural
starting position, such as the one in Figure~\ref{10x10 checkers starting},
remains open.
On the other hand, deciding the outcome of an arbitrary configuration is
PSPACE-hard \cite{Fraenkel-Garey-Johnson-Schaefer-Yesha-1978}.  If a polynomial
bound is placed on the number of moves that are allowed in between jumps (which
is a reasonable generalization of the drawing rule in standard Checkers
\cite{Fraenkel-Garey-Johnson-Schaefer-Yesha-1978}), then the problem is in
PSPACE and hence is PSPACE-complete.  Without such a restriction, however,
Checkers is EXPTIME-complete \cite{Robson-1984}.

On the other hand, certain simple questions about Checkers can be answered in
polynomial time
\cite{Fraenkel-Garey-Johnson-Schaefer-Yesha-1978,Demaine-Demaine-Eppstein-2002}.
Can one player remove all the other player's pieces in one move (by several
jumps)?  Can one player king a piece in one move?  Because of the notion of
parity on $n \times n$ boards, these questions reduce to checking the existence
of an Eulerian path or general path, respectively, in a particular directed
graph; see
\cite{Fraenkel-Garey-Johnson-Schaefer-Yesha-1978,Demaine-Demaine-Eppstein-2002}.
However, for boards defined by general graphs, at least the first question
becomes NP-complete \cite{Fraenkel-Garey-Johnson-Schaefer-Yesha-1978}.

\subsection{Go}

Presented at the same conference as the Checkers result in the previous section
(FOCS'78), Lichtenstein and Sipser \cite{Lichtenstein-Sipser-1980} proved that
the classic Asian game of Go is also PSPACE-hard for an arbitrary
configuration on an $n \times n$ board.
Go has few rules: (1) players take turns either passing or placing stones of
their color on positions on the board; (2) if a new black stone (say) causes a
collection of white stones to be completely surrounded by black stones, the
white stones are removed; and (3) a ko rule preventing repeated configurations.
Depending on the country, there are several variations of the ko rule; see
\cite{Berlekamp-Wolfe-1994}.  Go does not follow normal play: the winner in Go
is the player with the highest score at the end of the game. A player's score is 
counted as either the number of stones of his color on the board plus empty
spaces surrounded by his stones (area counting), or as  empty spaces surrounded
by his stones plus captured stones (territory counting), again
varying by country.

\begin{wrapfigure}{r}{2.5in}
  \centering
  \includegraphics[width=2.5in]{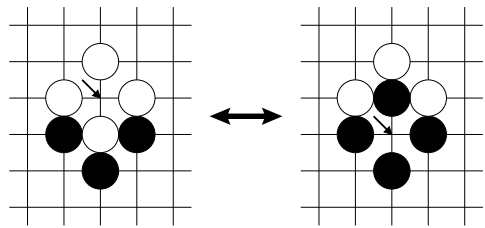}
  \caption{A simple form of ko in Go.}
  \label{simple ko}
\end{wrapfigure}

The PSPACE-hardness proof of Lichtenstein and Sipser
\cite{Lichtenstein-Sipser-1980} does not involve any situations called kos,
where the ko rule must be invoked to avoid infinite play.
In contrast, Robson
\cite{Robson-1983} proved that Go is EXPTIME-complete under Japanese rules
when kos are involved,
and indeed used judiciously.  The type of ko used in this reduction is shown in
Figure~\ref{simple ko}.  When one of the players makes a move shown in the
figure, the ko rule prevents (in particular) the other move shown in the figure
to be made immediately afterwards.

Robson's proof relies on properties of the Japanese rules for both the upper and
lower bounds. For other rulesets, all that is known is that Go is PSPACE-hard and
in EXPSPACE. In particular, the ``superko'' variant of the ko rule (as used 
in, e.g., the U.S.A.\ and New Zealand), which prohibits recreation of any former board
position, suggests EXPSPACE-hardness, by a result of Robson for no-repeat games \cite{Robson-1984a}.
However, if all dynamical state in the game occurs in kos, as it does in the 
EXPTIME-hardness construction, then the game is still in EXPTIME, because
then it is an instance of Undirected Vertex Geography (Section~\ref{More Games on Graphs}),
which can be solved in time polynomial in the graph size.
(In this case the graph is all the possible game positions, of which there are
exponentially  many.)

There are also several results for more restricted Go positions.
Wolfe \cite{Wolfe-2002} shows that even Go endgames are
PSPACE-hard.  More precisely, a \emph{Go endgame} is when the game has reduced
to a sum of Go subgames, each equal to a polynomial-size game tree.
This proof is based on several connections between Go and combinatorial game
theory detailed in a book by Berlekamp and Wolfe \cite{Berlekamp-Wolfe-1994}.
Cr\^{a}\c{s}maru and Tromp \cite{Crasmaru-Tromp-2000} show that it is
PSPACE-complete to determine whether a ladder (a repeated pattern of capture
threats) results in a capture.
Finally, Cr\^{a}\c{s}maru \cite{Crasmaru-1999} shows that it is NP-complete
to determine the status of certain restricted forms of life-and-death problems
in Go.

\subsection{Five-in-a-Row (Gobang)}

\emph{Five-in-a-Row} or \emph{Gobang}
\cite[pp.~738--740]{Berlekamp-Conway-Guy-2001} is another game on a Go board
in which players take turns placing a stone of their color.  Now the goal of
the players is to place at least $5$ stones of their color in a row either
horizontally, vertically, or diagonally.  This game is similar to Go-Moku
\cite[p.~740]{Berlekamp-Conway-Guy-2001}, which does not count $6$ or more
stones in a row, and imposes additional constraints on moves.

Reisch \cite{Reisch-1980} proved that deciding the outcome of a Gobang position
is PSPACE-complete.  He also observed that the reduction can be adapted to the
rules of $k$-in-a-Row for fixed $k$.  Although he did not specify exactly which
values of $k$ are allowed, the reduction would appear to generalize to any $k
\geq 5$.

\subsection{Chess}

Fraenkel and Lichtenstein \cite{Fraenkel-Lichtenstein-1981} proved that a
generalization of the classic game Chess to $n \times n$ boards is
EXPTIME-complete.  Specifically, their generalization has a unique
king of each color, and for each color the numbers of pawns,
bishops, rooks, and queens increase as some fractional power of $n$.
(Knights are not needed.)
The initial configuration is unspecified; what is EXPTIME-hard is to determine
the winner (who can checkmate) from an arbitrary specified configuration.

\subsection{Shogi}
\label{Shogi}

\emph{Shogi} is a Japanese game along lines similar to Chess, but with rules
too complex to state here.  Adachi, Kamekawa, and Iwata
\cite{Adachi-Kamekawa-Iwata-1987} proved that deciding the outcome of a Shogi
position is EXPTIME-complete.
Recently,
Yokota et al.\ \cite{Yokota-Tsukiji-Kitagawa-Morohashi-Iwata-2001}
proved that a more restricted form of Shogi, \emph{Tsume-Shogi}, in which
the first player must continually make oh-te (the equivalent of check in
Chess), is also EXPTIME-complete.

\begin{wrapfigure}{r}{1.5in}
  \centering
  \vspace*{-\intextsep}
  \includegraphics[width=1.5in]{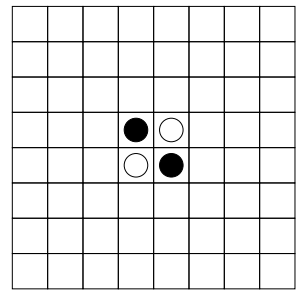}
  \caption{Initial configuration in Othello.}
  \label{othello starting}
\end{wrapfigure}

\subsection{Othello (Reversi)}

\emph{Othello} (\emph{Reversi}) is a classic game on an $8 \times 8$ board,
starting from the initial configuration shown in Figure~\ref{othello starting},
in which players alternately place pieces of their color in unoccupied squares.
Moves are restricted to cause, in at least one row, column, or diagonal, a
consecutive sequence of pieces of the opposite color to be enclosed by two
pieces of the current player's color.  As a result of the move, the enclosed
pieces ``flip'' color into the current player's color.
The winner is the player with the most pieces of their color when the
board is filled.

Generalized to an $n \times n$ board with an arbitrary initial configuration,
the game is clearly in PSPACE because only $n^2-4$ moves can be made.
Furthermore, Iwata and Kasai \cite{Iwata-Kasai-1994} proved that the game is
PSPACE-complete.

\subsection{Hackenbush}

\emph{Hackenbush} is one of the standard examples of a combinatorial game in
\emph{Winning Ways}; see, e.g., \cite[pp.~1--6]{Berlekamp-Conway-Guy-2001}.  A
position is given by a graph with each edge colored either red (Left), blue
(Right), or green (neutral), and with certain vertices marked as \emph{rooted}.
Players take turns removing an edge of an appropriate color (either neutral or
their own color), which also causes all edges not connected to a rooted vertex
to be removed.  The winner is determined by normal play.

Chapter 7 of \emph{Winning Ways} \cite[pp.~189--227]{Berlekamp-Conway-Guy-2001}
proves that determining the \emph{value} of a red-blue Hackenbush position is
NP-hard.
The reduction is from minimum Steiner tree in graphs.  It applies to
a restricted form of hackenbush positions, called \emph{redwood beds},
consisting of a red bipartite graph, with each vertex on one side attached to a
red edge, whose other end is attached to a blue edge, whose other end is
rooted.

\subsection{Domineering (Crosscram) and Cram}
\label{Domineering}

\emph{Domineering} or \emph{crosscram}
\cite[pp.~119--126]{Berlekamp-Conway-Guy-2001} is a partizan game
involving placement of horizontal and vertical dominoes in a grid;
a typical starting position is an $m \times n$ rectangle.
Left can play only vertical dominoes and Right can play only horizontal
dominoes, and dominoes must remain disjoint.  The winner is determined by
normal play.

The complexity of Domineering, computing either the outcome or the value of a
position, remains open.  Lachmann, Moore, and Rapaport
\cite{Lachmann-Moore-Rapaport-2000} have shown that the winner and a winning
strategy can be computed in polynomial time for $m \in \{1,2,3,4,5,7,9,11\}$
and all $n$.  These algorithms do not compute the value of the game, nor
the optimal strategy, only a winning strategy.

\emph{Cram} \cite{Gardner-1986-cram},
\cite[pp.~502--506]{Berlekamp-Conway-Guy-2001} is the impartial version of
Domineering in which both players can place horizontal and vertical dominoes.
The outcome of Cram is easy to determine for rectangles having an even number
of squares \cite{Gardner-1986-cram}: if both sides are even, the second player
can win by a symmetry strategy (reflecting the first player's move through both
axes); and if precisely one side is even, the first player can win by playing
the middle two squares and then applying the symmetry strategy.
It seems open to determine the outcome for a rectangle having two odd sides.
The complexity of Cram for general boards also remains open.

\emph{Linear Cram} is Cram in a $1 \times n$ rectangle, where the game quickly
splits into a sum of games.  This game can be solved easily by applying the
Sprague-Grundy theory and dynamic programming; in fact, there is a simpler
solution based on proving that its behavior is periodic in $n$
\cite{Guy-Smith-1956}.
The variation on Linear Cram in which $1 \times k$ rectangles are placed
instead of dominoes can also be solved via dynamic programming, but whether the
behavior is periodic remains open even for $k=3$ \cite{Guy-Smith-1956}.
Mis\`ere Linear Cram also remains unsolved \cite{Gardner-1986-cram}.


\subsection{Dots-and-Boxes, Strings-and-Coins, and Nimstring}
\label{Dots and Boxes}

\emph{Dots-and-Boxes} is a well-known children's game in which players take
turns drawing horizontal and vertical edges connecting pairs of dots in an
$m \times n$ subset of the lattice.
Whenever a player makes a move that encloses a unit square with drawn
edges, the player is awarded a point and must then draw another edge in the
same move.  The winner is the player with the most points when the entire grid
has been drawn.
Most of this section is based on Chapter 16 of \emph{Winning Ways}
\cite[pp.~541--584]{Berlekamp-Conway-Guy-2001}; another good reference is
a recent book by Berlekamp \cite{Berlekamp-2000}.


Gameplay in Dots-and-Boxes typically divides into two phases: the
\emph{opening} during which no boxes are enclosed, and the \emph{endgame}
during which boxes are enclosed in nearly every move; see Figure~\ref{dots
boxes endgame}.  In the endgame, the
``free move'' awarded by enclosing a square often leads to several squares
enclosed in a single move, following a \emph{chain}; see Figure~\ref{dots boxes
chain}.  Most children apply the greedy algorithm of taking the most squares
possible, and thus play entire chains of squares.  However, this strategy
forces the player to open another chain (in the endgame).  A simple improved
strategy is called \emph{double dealing}, which forfeits the last two squares
of the chain, but forces the opponent to open the next chain.
The double-dealer is said to \emph{remain in control};
if there are long-enough chains,
this player will win (see \cite[p.~543]{Berlekamp-Conway-Guy-2001}
for a formalization of this statement).

\begin{figure}[t]\hfil
\begin{minipage}{1.5in}
  \centering
  \includegraphics[width=1.5in]{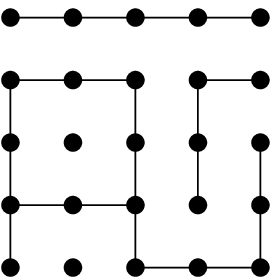}
  \caption{A Dots-and-Boxes endgame.}
  \label{dots boxes endgame}
\end{minipage}\hfil\hfil
\begin{minipage}{3in}
  \centering
  \ifpdf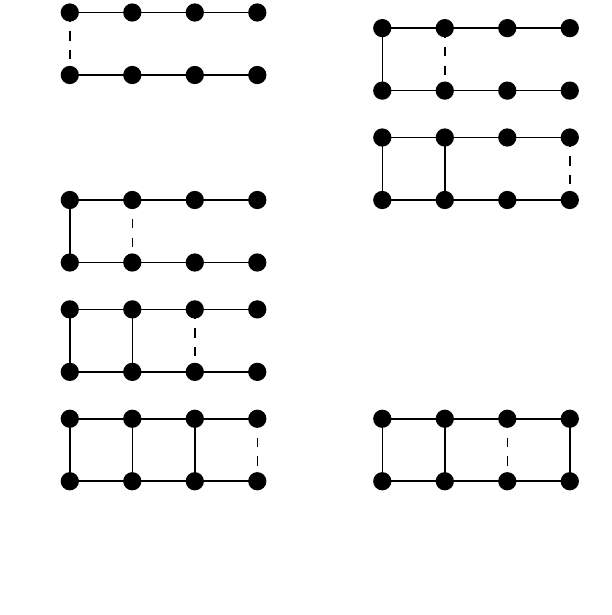\else\input{dots_boxes_chain.pstex_t}\fi
  \caption{Chains and double-dealing in Dots-and-Boxes.}
  \label{dots boxes chain}
\end{minipage}\hfil
\end{figure}

A generalization arising from the dual of Dots-and-Boxes is
\emph{Strings-and-Coins} \cite[pp.~550--551]{Berlekamp-Conway-Guy-2001}.
This game involves a sort of graph whose vertices
are \emph{coins} and whose edges are \emph{strings}.  The coins may be tied to
each other and to the ``ground'' by strings; the latter connection can be
modeled as a loop in the graph.  Players alternate cutting strings (removing
edges), and if a coin is thereby freed, that player collects the coin and cuts
another string in the same move.  The player to collect the most coins wins.

Another game closely related to Dots-and-Boxes is \emph{Nimstring}
\cite[pp.~552--554]{Berlekamp-Conway-Guy-2001},
which has the same rules as Strings-and-Coins,
except that the winner is determined by normal play.
Nimstring is in fact a special case of Strings-and-Coins
\cite[p.~552]{Berlekamp-Conway-Guy-2001}: if we add a chain of more than $n+1$
coins to an instance of Nimstring having $n$ coins, then ideal play of the
resulting string-and-coins instance will avoid opening the long chain for as
long as possible, and thus the player to move last in the Nimstring instance
wins string and coins.

\emph{Winning Ways} \cite[pp.~577--578]{Berlekamp-Conway-Guy-2001} argues that
Strings-and-Coins is NP-hard as follows.  Suppose that you have gathered
several coins but your opponent gains control.
Now you are forced to lose the Nimstring game, but given your initial lead, you
still may win the Strings-and-Coins game.  Minimizing the number of coins lost
while your opponent maintains control is equivalent to finding the maximum
number of vertex-disjoint cycles in the graph, basically because the equivalent
of a double-deal to maintain control once an (isolated) cycle is opened results
in forfeiting four squares instead of two.  We observe that by making the
difference between the initial lead and the forfeited coins very small (either
$-1$ or $1$), the opponent also cannot win by yielding control.  Because the
cycle-packing problem is NP-hard on general graphs, determining the outcome of
such string-and-coins endgames is NP-hard.  Eppstein \cite{Eppstein-cgt-hard}
observes that this reduction should also apply to endgame instances of
Dots-and-Boxes by restricting to maximum-degree-three planar graphs.
Embeddability of such graphs in the square grid follows because long chains and
cycles (longer than two edges for chains and three edges for cycles) can be
replaced by even longer chains or cycles
\cite[p.~561]{Berlekamp-Conway-Guy-2001}.


It remains open whether Dots-and-Boxes or Strings-and-Coins are in NP
or PSPACE-complete from an arbitrary configuration.
Even the case of a $1 \times n$ grid of boxes is
not fully understood from a Combinatorial Game Theory perspective
\cite{Guy-Nowakowski-2002}.

\begin{wrapfigure}{r}{3in}
  \centering
  \vspace*{-\intextsep}\vspace*{-2ex}
  \includegraphics[width=3in]{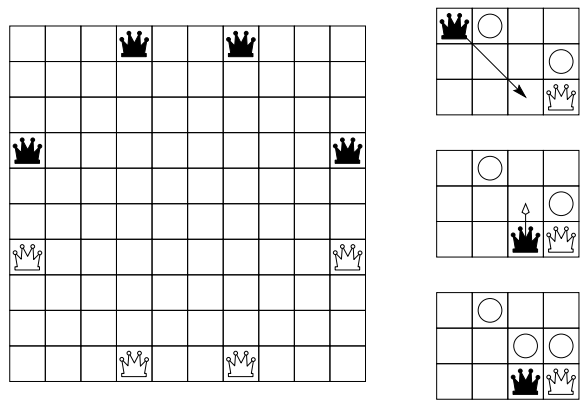}
  \caption{The initial position in Amazons (left) and an example of
    black trapping a white amazon (right).}
  \label{amazons examples}
\end{wrapfigure}

\subsection{Amazons}

\emph{Amazons} is a game invented by Walter Zamkauskas in 1988, containing
elements of Chess and Go.  Gameplay takes place on a $10 \times 10$ board with
four \emph{amazons} of each color arranged as in Figure~\ref{amazons examples}
(left).  In each turn, Left [Right] moves a black [white] amazon to any
unoccupied square accessible by a Chess queen's move, and fires an arrow to any
unoccupied square reachable by a Chess queen's move from the amazon's new
position.  The arrow (drawn as a circle) now occupies its square; amazons and
shots can no longer pass over or land on this square.  The winner is determined
by normal play.

Gameplay in Amazons typically split into a sum of simpler games because arrows
partition the board into multiple components.  In particular, the
\emph{endgame} begins when each component of the game contains amazons of only
a single color.  Then the goal of each player is simply to maximize the number
of moves in each component.  Buro \cite{Buro-2000} proved that maximizing the
number of moves in a single component is NP-complete (for $n \times n$ boards).
In a general endgame, deciding the outcome may not be in NP because it is
difficult to prove that the opponent has no better strategy.  However, Buro
\cite{Buro-2000} proved that this problem is \emph{NP-equivalent}
\cite{Garey-Johnson-1979}, that is, the problem can be solved by a polynomial
number of calls to an algorithm for any NP-complete problem, and vice versa.

Like Conway's Angel Problem (Section \ref{Conway's Angel Problem}), the complexity of 
deciding the outcome of a general Amazons position remained open for several
years, only to be solved nearly simultaneously by multiple people.
Furtak, Kiyomi, Takeaki, and Buro \cite{Furtak-Kiyomi-Takeaki-Buro-2005} give two
independent proofs of PSPACE-completeness: one a reduction from Hex, and the other
a reduction from Vertex Geography. The latter reduction applies even for positions
containing only a single black and a single white amazon. Independently,
Hearn \cite{Hearn-2005, Hearn-2006, Hearn-2008} gave a Constraint Logic reduction showing
PSPACE-completeness.

\begin{wrapfigure}{r}{1.8in}
  \centering
  \vspace*{-\intextsep}
  \vspace*{-8ex}
  \includegraphics[width=1.5in]{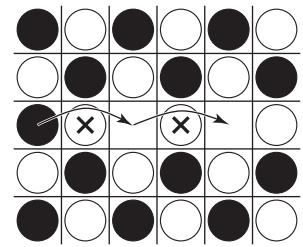}
  \caption{One move in Konane consisting of two jumps.}
  \vspace*{-2ex}
  \label{konane example}
\end{wrapfigure}

\subsection{Konane}
\label{Konane}

Konane, or Hawaiian Checkers, is a game that has been played in Hawaii since
preliterate times.  Konane is played on a rectangular board (typically ranging
in size from $8 \times 8$ to $13 \times 20$) which is initially filled with
black and white stones in a checkerboard pattern.  To begin the game, two
adjacent stones in the middle of the board or in a corner are removed.
Then, the players alternate making moves.  Moves are made as in Peg Solitaire
(Section~\ref{Peg Solitaire}); indeed, Konane may be thought of as a kind
of two-player peg solitaire.  A player moves a stone of his color by jumping
it over a horizontally or vertically adjacent stone of he opposite color,
into an empty space.  (See Figure~\ref{konane example}.)  Jumped stones are
captured and removed from play.  A stone may make multiple successive jumps
in a single move, so long as they are in a straight line; no turns are allowed
within a single move.  The first player unable to move wins. 

Hearn proved that Konane is PSPACE-complete \cite{Hearn-2006, Hearn-2008} by 
a reduction from Constraint Logic.
There have been some positive results for restricted configurations.
Ernst \cite{Ernst-95} derives Combinatorial-Game-Theoretic values for
several interesting positions.
Chan and Tsai \cite{Chan-Tsai-2002} analyze the $1 \times n$ game,
but even this version of the game is not yet solved.

\subsection{Phutball}
\label{Phutball}

\begin{figure}
  \centering
  \includegraphics[width=\linewidth]{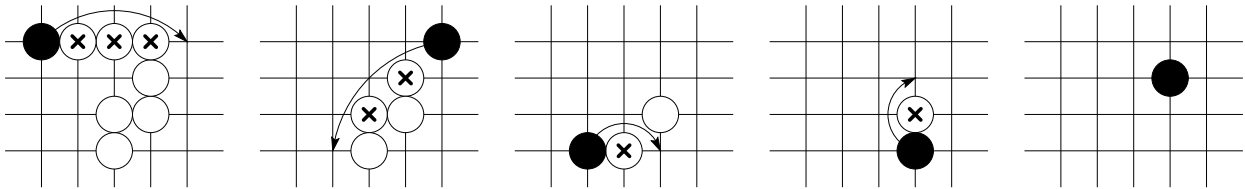}
  \caption{A single move in Phutball consisting of four jumps.}
  \label{phutball example}
\end{figure}

Conway's game of \emph{Philosopher's Football} or \emph{Phutball}
\cite[pp.~752--755]{Berlekamp-Conway-Guy-2001} involves white and black stones
on a rectangular grid such as a Go board.  Initially, the unique black stone
(the \emph{ball}) is placed in the middle of the board, and there are no white
stones.  Players take turns either placing a white stone in any unoccupied
position, or moving the ball by a sequence of \emph{jumps} over consecutive
sequences of white stones each arranged horizontally, vertically, or
diagonally.  See Figure~\ref{phutball example}.  A jump causes immediate
removal of the white stones jumped over, so those stones cannot be used for a
future jump in the same move.  Left and Right have opposite sides of the grid
marked as their \emph{goal lines}.  Left's goal is to end a move with the ball
on or beyond Right's goal line, and symmetrically for Right.

Phutball is inherently loopy and it is not clear that either player has a
winning strategy: the game may always be drawn out indefinitely.  One
counterintuitive aspect of the game is that white stones placed by one player
may be ``corrupted'' for better use by the other player.  Recently, however,
Demaine, Demaine, and Eppstein \cite{Demaine-Demaine-Eppstein-2002} found an
aspect of Phutball that could be analyzed.  Specifically, they proved that
determining whether the current player can win in a single move (``mate in 1''
in Chess) is NP-complete.  This result leaves open the complexity of
determining the outcome of a given game position.

\subsection{Conway's Angel Problem}
\label{Conway's Angel Problem}

A formerly long-standing open problem was Conway's \emph{Angel Problem}
\cite{Berlekamp-Conway-Guy-2001}.
Two players, the Angel and the Devil, alternate play on an
infinite square grid.
The Angel can move to any valid position within $k$ horizontal distance
and $k$ vertical distance from its present position.
The Devil can teleport to an arbitrary square other than where the Angel is
and ``eat'' that square, preventing the Angel from landing on
(but not leaping over) that square in the future.
The Devil's goal is to prevent the Angel from moving.

It was long known that an Angel of power $k=1$ can be stopped
\cite{Berlekamp-Conway-Guy-2001},
so the Devil wins,
but the Angel was not known to be able to escape for any $k>1$.
(In the original open problem statement, $k = 1000$.)
Recently, four independent proofs established that a sufficiently strong Angel
can move forever, securing the Angel as the winner.
M\'ath\'e \cite{Mathe-2007} and Kloster \cite{Kloster-2007}
showed that $k = 2$ suffices; Bowditch \cite{Bowditch-2007}
showed that $k = 4$ suffices; and G\'{a}cs \cite{Gacs-2007}
showed that some $k$ suffices.
In particular, Kloster's proof gives an explicit algorithmic
winning strategy for the $k=2$ Angel.

\subsection{Jenga}

\emph{Jenga} is a popular stacked-block game invented by Leslie Scott in the
1970s and now marketed by Hasbro.  Two players alternate moving individual
blocks in a tower of blocks, and the first player to topple the tower
(or cause any additional blocks to fall) loses.
Each block is $1 \times 1 \times 3$ and lies horizontally.
The initial $3 \times 3 \times n$ tower alternates levels of three blocks each,
so that blocks in adjacent levels are orthogonal.
(In the commercial game, $n=18$.)
In each move, the player removes any block that is below the topmost complete
($3$-block) level, then places that block in the topmost level
(starting a new level if the existing topmost level is complete),
orthogonal to the blocks in the (complete) level below.
The player loses if the tower becomes instable, that is, the center of gravity
of the top $k$ levels projects outside the convex hull of the contact area
between the $k$th and $(k+1)$st layer.

Zwick \cite{Zwick-2002-Jenga} proved that the physical stability condition
of Jenga can be restated combinatorially simply by constraining allowable
patterns on each level and the topmost three levels.  Specifically, write a
$3$-bit vector to specify which blocks are present in each level.
Then a tower is stable if and only if no level except possibly the top
is $100$ or $001$ and the three topmost levels from bottom to top are none of
$010,010,100$; $010,010,001$; $011,010,100$; or $110,010,001$.
Using this characterization, Zwick proves that the first player wins
from the initial configuration if and only if $n=2$ or $n \geq 4$ and
$n \equiv 1$ or $2 \pmod 3$,
and gives a simple characterization of winning moves.
It remains open whether such an efficient solution can be obtained
in the generalization to odd numbers $k > 3$ of blocks in each level.
(The case of even $k$ is a second-player win by a simple mirror strategy.)

\section{Algorithms for Puzzles}
\label{Algorithms for Puzzles}

Many puzzles (one-player games) have short solutions and are NP-complete.
However, several puzzles based on motion-planning problems are harder,
PSPACE-hard.  Usually such puzzles occupy a bounded board or region,
so they are also PSPACE-complete.  A common method to
prove that such puzzles are in PSPACE is to give a simple low-space
nondeterministic algorithm that guesses the solution, and apply Savitch's
theorem \cite{Savitch-1970} that PSPACE = NPSPACE (nondeterministic polynomial
space).
However, when generalized to the entire plane and unboundedly many
pieces, puzzles often become undecidable.

This section briefly surveys some of these results, following the structure
of the previous section.

\subsection{Instant Insanity}

Given $n$ cubes, each face colored one of $n$ colors, is it possible to stack
the cubes so that each color appears exactly once on each of the $4$ sides of
the stack?  The case of $n=4$ is a puzzle called Instant Insanity distributed
by Parker Bros.  In one of the first papers on hardness of puzzles and games
people play, Robertson and Munro \cite{Robertson-Munro-1978} proved that
this \emph{generalized Instant Insanity} problem is NP-complete.

The \emph{cube stacking game} is a two-player game based on this puzzle.  Given
an ordered list of cubes, the players take turns adding the next cube to the
top of the stack with a chosen orientation.  The loser is the first player to
add a cube that causes one of the four sides of the stack to have a color
repeated more than once.  Robertson and Munro \cite{Robertson-Munro-1978}
proved that this game is PSPACE-complete, intended as a general illustration
that NP-complete puzzles tend to lead to PSPACE-complete games.

\subsection{Cryptarithms (Alphametics, Verbal Arithmetic)}

\emph{Cryptarithms} or \emph{alphametics} or \emph{verbal arithmetic}
are classic puzzles involving an equation of symbols,
the original being Dudeney's $SEND+MORE=MONEY$ from 1924
\cite{Dudeney-1924-cryptarithm}, in which each symbol (e.g., $M$)
represents a consistent digit (between $0$ and $9$).
The goal is to determine an assignment of digits to symbols that satisfies the
equation.  Such problems can easily be solved in polynomial time by enumerating
all $10!$ assignments.  However, Eppstein \cite{Eppstein-1987} proved that it
is NP-complete to solve the generalization to base $\Theta(n^3)$ (instead of
decimal) and $\Theta(n)$ symbols (instead of~$26$).

\subsection{Crossword Puzzles and Scrabble}
\label{Crossword Puzzles and Scrabble}

Perhaps one of the most popular puzzles are \emph{crossword puzzles},
going back to 1913 and today appearing in almost every newspaper,
and the subject of the recent documentary \emph{Wordplay} (2006).
Here it is easiest to model the problem of designing crossword puzzles,
ignoring the nonmathematical notion of clues.  Given a list of words
(the dictionary), and a rectangular grid with some squares obstacles and others
blank, can we place a subset of the words into horizontally or vertically
maximal blank strips so that crossing words have matching letters?
Lewis and Papadimitriou \cite[p.~258]{Garey-Johnson-1979}
proved that this question is NP-complete, even when the grid has no
obstacles so every row and column must form a word.

Alternatively, this problem can be viewed as the ultimate form of
crossword puzzle solving, without clues.  In this case it would be interesting
to know whether the problem remains NP-hard even if every word in the given
list must be used exactly once, so that the single clue could be
``use these words''.
A related open problem is \emph{Scrabble}, which we are not aware of
having been studied.
The most natural theoretical question is perhaps the one-move version:
given the pieces in hand (with letters and scores), and given
the current board configuration (with played pieces and
available double/triple letter/word squares), what move maximizes score?
Presumably the decision question is NP-complete.
Also open is the complexity of the two-player game, say in the
perfect-information variation where both players know the sequence in
which remaining pieces will be drawn as well as the pieces in the
opponent's hand.
Presumably determining a winning move from a given position in this game
is PSPACE-complete.


%
%
%
%
%

\subsection{Pencil-and-Paper Puzzles: Sudoku and Friends}
\label{Pencil-and-Paper}

\emph{Sudoku} or \emph{Number Place} is a pencil-and-paper puzzle that became
popular worldwide starting around 2005
\cite{Delahaye-2006,Hayes-2006-Sudoku}.
American architect Howard Garns first published the puzzle
in the May 1979 (and many subsequent)
\emph{Dell Pencil Puzzles and Word Games} (without a byline);
then Japanese magazine \emph{Monthly Nikolist} imported the puzzle
in 1984, trademarking the name Sudoku (single numbers);
then the idea spread throughout Japanese publications;
finally Wayne Gould published his own computer-generated puzzles in
\emph{The Times} in 2004, shortly after which many newspapers and magazines
adopted the puzzle.
The usual puzzle consists of an $9 \times 9$ grid of squares,
divided into a $3 \times 3$ arrangement of $3 \times 3$ tiles.
Some grid squares are initially filled with digits between $1$ and $9$,
and some are blank.  The goal is to fill the blank squares so that
every row, column, and tile has all nine digits without repetition.

Sudoku naturally generalizes to an $n^2 \times n^2$ grid of squares,
divided into an $n \times n$ arrangement of $n \times n$ tiles.
Yato and Seta \cite{Yato-Seta-2003,Yato-2003}
proved that this generalization is NP-complete.
In fact, they proved a stronger completeness result, in the class of
Another Solution Problems (ASP), where one is given one or more solutions
and wishes to find another solution.
Thus, in particular, given a Sudoku puzzle and an intended solution,
it is NP-complete to determine whether there is another solution,
a problem arising in puzzle design.
Most Sudoku puzzles give the promise that they have a unique solution.
Valiant and Vazirani \cite{Valiant-Vazirani-1986} proved that adding such
a uniqueness promise keeps a problem NP-hard under randomized reductions,
so there is no polynomial-time solution to uniquely solvable Sudokus
unless RP~$=$~NP.


ASP-completeness (in particular, NP-completeness) has been established for
six other paper-and-pencil puzzles by Japanese publisher Nikoli:
Nonograms, Slitherlink, Cross Sum, Fillomino, Light Up, and LITS.
In a \emph{Nonogram} or \emph{Paint by Numbers} puzzle
\cite{Ueda-Nagao-1996},
we are given a sequence of integers on each row and column of a rectangular
matrix, and the goal is to fill in a subset of the squares in the matrix
so that, in each row and column, the maximal contiguous runs of filled squares
have lengths that match the specified sequence.
In \emph{Slitherlink} \cite{Yato-Seta-2003,Yato-2003},
we are given labels between $0$ and $4$ on some subset of faces
in a rectangular grid, and the goal is to draw a simple cycle on the grid
so that each labeled face is surrounded by the specified number of edges.
In \emph{Kakuro} or \emph{Cross Sum} \cite{Yato-Seta-2003},
we are given a polyomino (a rectangular grid where only some squares
may be used), and an integer for each maximal contiguous
(horizontal or vertical) strip of squares,
and the goal is to fill each square with a digit between $1$ and $9$
such that each strip has the specified sum and has no repeated digit.
In \emph{Fillomino} \cite{Yato-2003},
we are given a rectangular grid in which some squares
have been filled with positive integers, and the goal is to fill the remaining
squares with positive integers
so that every maximal connected region of equally numbered squares consists
of exactly that number of squares.
In \emph{Light Up} (\emph{Akari}) \cite{McPhail-2005-LightUp,McPhail-2007-LITS},
we are given a rectangular grid in which squares are either rooms or walls
and some walls have a specified integer between $0$ and $4$,
and the goal is to place lights in a subset of the rooms such that
each numbered wall has exactly the specified number of (horizontally or
vertically) adjacent lights, every room is horizontally or vertically visible
from a light, and no two lights are horizontally or vertically visible
from each other.
In \emph{LITS} \cite{McPhail-2007-LITS}, we are given a division of a
rectangle into polyomino pieces, and the goal is to choose a tetromino
(connected subset of four squares) in each polyomino such that
the union of tetrominoes is connected yet induces no $2 \times 2$ square.
As with Sudoku, it is NP-complete to both find solutions and test uniqueness
of known solutions in all of these puzzles.

NP-completeness has been established for seven other pencil-and-paper games
published by Nikoli: Tentai Show, Masyu, Bag, Nurikabe, Hiroimono, Heyawake,
and Hitori.
In \emph{Tentai Show} or \emph{Spiral Galaxies} \cite{Friedman-2002-spiral},
we are given a rectangular grid with dots at some vertices, edge midpoints,
and face centroids, and the goal is to divide the rectangle into
exactly one polyomino piece per dot that is two-fold rotationally symmetric
around the dot.
In \emph{Masyu} or \emph{Pearl Puzzles} \cite{Friedman-2002-pearl},
we are given a rectangular grid with some squares containing white or black
pearls, and the goal is to find a simple path through the squares that visits
every pearl, turns $90^\circ$ at every black pearl, does not turn immediately
before or after black pearls, goes straight through every white pearl,
and turns $90^\circ$ immediately before or after every white pearl.
In \emph{Bag} or \emph{Corral Puzzles} \cite{Friedman-2002-corral},
we are given a rectangular grid with some squares labeled with positive
integers, and the goal is to find a simple cycle on the grid that encloses
all labels and such that the number of squares horizontally and vertically
visible from each labeled square equals the label.
In \emph{Nurikabe} \cite{McPhail-2003,Holzer-Klein-Kutrib-2004},
we are given a rectangular grid with some squares labeled with positive
integers, and the goal is to find a connected subset of unlabeled squares that
induces no $2 \times 2$ square and whole removal results in exactly one region
per labeled square whose size equals that label.
McPhail's reduction \cite{McPhail-2003} uses labels $1$ through $5$,
while Holzer et al.'s reduction \cite{Holzer-Klein-Kutrib-2004} only uses
labels $1$ and $2$ (just $1$ would be trivial) and works without
the connectivity rule and/or the $2 \times 2$ rule.
In \emph{Hiroimono} or \emph{Goishi Hiroi} \cite{Andersson-2007},
we are given a collection of stones at vertices of a rectangular grid,
and the goal is to find a path that visits all stones, changes directions
by $\pm 90^\circ$ and only at stones, and removes stones as they are visited
(similar to Phutball in Section~\ref{Phutball}).
In \emph{Heyawake} \cite{Holzer-Ruepp-2007},
we are given a subdivision of a rectangular grid into rectangular rooms,
some of which are labeled with a positive integer, and the goal is to
paint a subset of unit squares so that the number of painted squares in
each labeled room equals the label, painted squares are never (horizontally
or vertically) adjacent, unpainted squares are connected (via horizontal
and vertical connections), and maximal contiguous (horizontal or vertical)
strips of squares intersect at most two rooms.
In \emph{Hitori} \cite{Hearn-2008-Hitori},
we are given a rectangular grid with each square labeled with an integer,
and the goal is to paint a subset of unit squares so that
every row and every column has no repeated unpainted label (similar to Sudoku),
painted squares are never (horizontally or vertically) adjacent,
and unpainted squares are connected (via horizontal and vertical connections).

A different kind of pencil-and-paper puzzle is \emph{Morpion Solitaire},
popular in several European countries.  The game starts with some 
configuration of points drawn at the intersections of a square grid
(usually in a standard cross pattern).
A move consists of placing a new point at a grid intersection,
and then drawing a horizontal, vertical, or diagonal line segment
connecting five consecutive points that include the new one.
Line segments with the same direction cannot share a point
(the \emph{disjoint model}); alternatively, line segments with the same
direction may overlap only at a common endpoint (the \emph{touching model}).
The goal is to maximize the number of moves before no moves are possible.
Demaine, Demaine, Langerman, and Langerman
\cite{Demaine-Demaine-Langerman-Langerman-2006}
consider this game generalized to moves connecting any number $k+1$
of points instead of just~$5$.
In addition to bounding the number of moves from the standard cross
configuration, they prove complexity results for the general case.
They show that, in both game models and for $k \geq 3$,
it is NP-hard to find the longest play from a given pattern of $n$ dots,
or even to approximate the longest play within $n^{1-\epsilon}$
for any $\epsilon > 0$.
For $k > 3$, the problem is in fact NP-complete.
For $k = 3$, it is open whether the problem is in~NP,
and for $k = 2$ it could even be in~P.

A final NP-completeness result for pencil-and-paper puzzles is the
Battleship puzzle.  This puzzle is a one-player perfect-information
variant on the classic two-player imperfect-information game,
\emph{Battleship}.
In \emph{Battleships} or \emph{Battleship Solitaire}
\cite{Sevenster-2004}, we are given a list of $1 \times k$ ships
for various values of~$k$; a rectangular grid with some squares
labeled as water, ship interior, ship end, or entire ($1 \times 1$) ship;
and the number of ship (nonwater) squares that should be in each row and
each column.  The goal is to complete the square labeling to place the
given ships in the grid while matching the specified number of ship squares
in each row and column.

Several other pencil-and-paper puzzles remain unstudied from a complexity
standpoint.  For example, Nikoli's English website%
\footnote{\protect\url{http://www.nikoli.co.jp/en/puzzles/}}
suggests
Hashiwokakero,
Kuromasu (Where is Black Cells),
Number Link, 
Ripple Effect,
Shikaku, and
Yajilin (Arrow Ring); and
Nikoli's Japanese website%
\footnote{\protect\url{http://www.nikoli.co.jp/ja/puzzles/}}
lists more.


\begin{wrapfigure}{r}{3in}
  \centering
  \ifpdf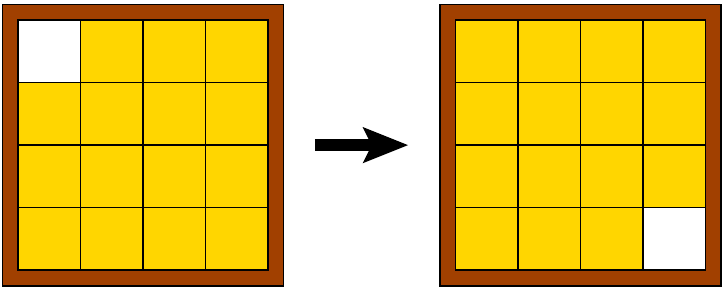\else\input{15_puzzle.pstex_t}\fi
  \caption{15 puzzle: Can you get from the left configuration to the right
    in 16 unit slides?}
  \label{15 puzzle}
\end{wrapfigure}

\subsection{Moving Tokens: Fifteen Puzzle and Generalizations}
\label{Moving Tokens}

The \emph{Fifteen Puzzle} or \emph{15 Puzzle}
\cite[p.~864]{Berlekamp-Conway-Guy-2001} is a classic
puzzle consisting of fifteen square blocks numbered $1$ through $15$
in a $4 \times 4$ grid; the remaining sixteenth square in the grid
is a hole which permits blocks to slide.  The goal is to
order the blocks to be increasing in English reading order.
The (six) hardest solvable positions require exactly 80 moves
\cite{Bruengger-Marzetta-Fukuda-Nievergelt-1999}.
Slocum and Sonneveld \cite{Slocum-Sonneveld-2006} recently uncovered
the history of this late 19th-century puzzle, which was well-hidden by
popularizer Sam Loyd since his claim of having invented it.

A natural generalization of the Fifteen Puzzle is the \emph{$n^2-1$ puzzle} on
an $n \times n$ grid.  It is easy to determine whether a configuration of the
$n^2-1$ puzzle can reach another: the two permutations of the block numbers (in
reading order) simply need to match in \emph{parity}, that is, whether the
number of inversions (out-of-order pairs) is even or odd.
See, e.g., \cite{Archer-1999,Story-1879,Wilson-1974}.
When the puzzle is solvable, the required numbers moves is $\Theta(n^3)$
in the worst case \cite{Parberry-1995}.
On the other hand, it is NP-complete to find a solution using the fewest
possible slides from a given configuration \cite{Ratner-Warmuth-1990}.
It is also NP-hard to approximate the fewest slides within an additive
constant, but there is a polynomial-time constant-factor approximation
\cite{Ratner-Warmuth-1990}.

\begin{wrapfigure}{r}{1.5in}
  \centering
  \vspace*{-\intextsep}
  \vspace{-1ex}
  \ifpdf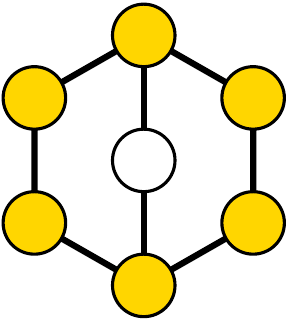\else\input{tricky_six_puzzle.pstex_t}\fi
  \caption{The Tricky Six Puzzle \protect\cite{Wilson-1974},
    \protect\cite[p.~868]{Berlekamp-Conway-Guy-2001}
    has six connected components of configurations.}
  \vspace{-1ex}
  \label{tricky six puzzle}
\end{wrapfigure}

The parity technique for determining solvability of the $n^2-1$ puzzle has been
generalized to a class of similar puzzles on graphs.  Consider an $N$-vertex
graph in which $N-1$ vertices have tokens labeled $1$ through $N-1$,
one vertex is empty (has no token), and each operation in the puzzle moves a
token to an adjacent empty vertex.
The goal is to reach one configuration from another.
This general puzzle encompasses the $n^2-1$ puzzle and several other puzzles
involving sliding balls in circular tracks, e.g., the Lucky Seven puzzle
\cite[p.~865]{Berlekamp-Conway-Guy-2001} or the puzzle shown in
Figure~\ref{tricky six puzzle}.  Wilson \cite{Wilson-1974},
\cite[p.~866]{Berlekamp-Conway-Guy-2001} characterized when these puzzles are
solvable, and furthermore characterized their group structure.  In most cases,
all puzzles are solvable (forming the symmetric group) unless the graph the
graph is bipartite, in which case half of the puzzles are solvable (forming the
alternating group).  In addition, there are three special situations: cycle
graphs, graphs having a cut vertex, and the special example in
Figure~\ref{tricky six puzzle}.

Even more generally, Kornhauser, Miller, and Spirakis
\cite{Kornhauser-Miller-Spirakis-1984} showed how to decide solvability
of puzzles with any number $k$ of labeled tokens on $N$ vertices.
They also prove that $O(N^3)$ moves always suffice,
and $\Omega(N^3)$ moves are sometimes necessary, in such puzzles.
Calinescu, Dumitrescu, and Pach \cite{Calinescu-Dumitrescu-Pach-2006}
consider the number of token ``shifts''---continuous moves along a path
of empty nodes---required in such puzzles.
They prove that finding the fewest-shift solution is NP-hard
in the infinite square grid and APX-hard in general graphs,
even if the tokens are unlabeled (identical).
On the positive side, they present a $3$-approximation for
unlabeled tokens in general graphs, an optimal solution for unlabeled tokens
in trees, an upper bound of $N$ slides for unlabeled tokens in general graphs,
and an upper bound of $O(N)$ slides for labeled tokens in the
infinite square grid.

Restricting the set of legal moves can make such puzzles harder.  Consider a graph with unlabeled
tokens on some vertices, and the constraint that the tokens must form an independent set on the 
graph (i.e., no two tokens are adjacent along an edge). A move is made by sliding a token along an
edge to an adjacent vertex, subject to maintaining the nonadjacency constraint. Then the problem of
determining whether a sequence of moves can ever move a given token, called \emph{Sliding Tokens} 
\cite{Hearn-Demaine-2005}, is PSPACE-complete.

\begin{wrapfigure}{r}{2in}
  \centering
  \vspace*{-\intextsep}
  \includegraphics[width=2in]{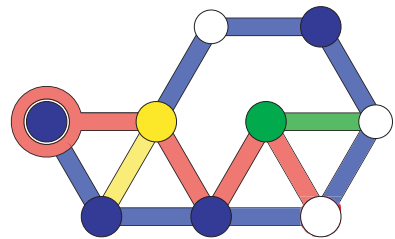}
  \caption{A Subway Shuffle puzzle with one red car, four blue cars,
    one yellow car, and one green car.  White nodes are empty.
    Moving the red car to the circled station requires 43 moves.}
  \label{subway example}
\end{wrapfigure}

\emph{Subway Shuffle} \cite{Hearn-2005a, Hearn-2006} is another constrained token-sliding puzzle on a graph.
In this puzzle both the tokens and the graph edges are colored; a move is to slide a token along an edge
of matching color to an unoccupied adjacent vertex. The goal is to move a specified token (the ``subway
car you have boarded'') to a specified vertex (your ``exit station''). A sample puzzle is shown in Figure~\ref{subway example}.
The complexity of determining whether there is a solution to a given puzzle is open. This open problem is quite fascinating:
solving the puzzle empirically seems hard, based on the rapid growth of minimum solution length with graph
size \cite{Hearn-2005a}. However, it is easy to determine whether a token may move at all by a sequence
of moves, evidently making the proof techniques used for Sliding Tokens and related problems useless for showing hardness.
Subway Shuffle can also be seen as a generalized version of $1 \times 1$ Rush Hour (Section~\ref{Sliding Blocks}).

Another kind of token-sliding puzzle is \emph{Atomix}, a computer game first
published in 1990.
Game play takes place on a rectangular board; pieces are either \emph{walls} (immovable
blocks) or \emph{atoms} of different types. A move is to slide an atom; in this case the atom
must slide in its direction of motion until it hits a wall (as in the PushPush family, below (Section~\ref{Pushing Blocks})).
The goal is to assemble a particular pattern of atoms (a molecule).
Huffner, Edelkamp, Fernau, and Niedermeier \cite{Huffner-Edelkamp-Fernau-Niedermeier-2001}
observed that Atomix is as hard as the $(n^2-1)$-puzzle, so it is NP-hard
to find a minimum-move solution.
Holzer and Schwoon \cite{Holzer-Schwoon-2004} later proved the stronger
result that it is PSPACE-complete to determine whether there is a solution.

\emph{Lunar Lockout} is another token-sliding puzzle, similar to Atomix in that
the tokens slide until stopped. Lunar Lockout was produced by ThinkFun at one time;
essentially the same game is now sold as ``Pete's Pike''. (Even earlier, the game was
called ``UFO''.)
In Lunar Lockout there are no walls or barriers; a token may only slide if there is another
token in place that will stop it. The goal is to get a particular token to a particular place.
Thus, the rules are fairly simple and natural; however, the complexity is open, though
there are partial results.
Hock \cite{Hock-2001} showed that Lunar Lockout is NP-hard, and that when the target token
may not revisit any position on the board, the problem becomes NP-complete.
Hartline and Libeskind-Hadas \cite{Hartline-Libeskind-Hadas-2003} show that a
generalization of Lunar Lockout which allows fixed blocks is PSPACE-complete.

\subsection{Rubik's Cube and Generalizations}

Alternatively, the $n^2-1$ puzzle can be viewed as a special case of
determining whether a permutation on $N$ items can be written as a product
(composition) of given generating permutations, and if so, finding such a
product.  This family of puzzles also includes \emph{Rubik's Cube}
(recently shown to be solvable in 26 moves \cite{Kunkle-Cooperman-2007})
and its many variations.
In general, the number of moves (terms) required to solve such a puzzle
can be exponential (unlike the Fifteen Puzzle).
Nonetheless, an $O(N^5)$-time algorithm can decide whether a given puzzle
of this type is solvable, and if so, find an implicit representation
of the solution \cite{Jerrum-1986}.
On the other hand, finding a solution with the fewest moves (terms) is
PSPACE-complete \cite{Jerrum-1985}.
When each given generator cyclically shifts just a bounded number of items,
as in the Fifteen Puzzle but not in a $k \times k \times k$ Rubik's Cube,
Driscoll and Furst \cite{Driscoll-Furst-1983} showed that such puzzles
can be solved in polynomial time using just $O(N^2)$ moves.
Furthermore, $\Theta(N^2)$ is the best possible bound in the worst case,
e.g., when the only permitted moves are swapping adjacent elements on a line.
See \cite{Kornhauser-Miller-Spirakis-1984,McKenzie-1984} for other
(not explicitly algorithmic) results on the maximum number of moves
for various special cases of such puzzles.

\subsection{Sliding Blocks and Rush Hour}
\label{Sliding Blocks}

\begin{wrapfigure}{r}{1.3in}
  \centering
  \vspace*{-\intextsep}
  \includegraphics{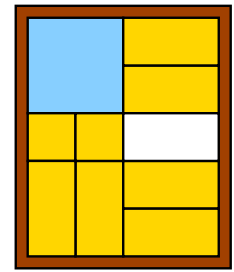}
  \caption{Dad's Puzzle \protect\cite{Gardner-1964}:
    moving the large square into the lower-left corner requires 59 moves.}
  \label{Dad's puzzle}
\end{wrapfigure}

A classic reference on a wide class of sliding-block puzzles is by Hordern
\cite{Hordern-1986}.  One general form of these puzzles is that rectangular
blocks are placed in a rectangular box, and each block can be moved
horizontally and vertically, provided the blocks remain disjoint.
The goal is usually either to move a particular block to a particular place,
or to re-arrange one configuration into another.
Figure~\ref{Dad's puzzle} shows an example which, according to Gardner
\cite{Gardner-1964}, may be the earliest (1909) and is the most widely sold
(after the Fifteen Puzzle, in each case).
Gardner \cite{Gardner-1964} first raised the question of whether there is an
efficient algorithm to solve such puzzles.
Spirakis and Yap \cite{Spirakis-Yap-1983} showed that achieving a specified
target configuration is NP-hard, and conjectured PSPACE-completeness.
Hopcroft, Schwartz, and Sharir \cite{Hopcroft-Schwartz-Sharir-1984}
proved PSPACE-completeness shortly afterwards, renaming the problem to
the ``Warehouseman's Problem''.
In the Warehouseman's Problem, there is no restriction on the sizes of blocks;
the blocks in the reduction grow with the size of the containing box.
By contrast, in most sliding-block puzzles,
the blocks are of small constant sizes.
Finally, Hearn and Demaine \cite{Hearn-Demaine-2002, Hearn-Demaine-2005}
showed that it is PSPACE-hard to decide whether a given piece can move at all
by a sequence of moves, even when all the blocks are $1 \times 2$ or
$2 \times 1$.
This result is best possible: the results above about unlabeled tokens in
graphs show that $1 \times 1$ blocks are easy to re-arrange.

A popular sliding-block puzzle is \emph{Rush Hour}, distributed by ThinkFun,
Inc.\ (formerly Binary Arts, Inc.).
We are given a configuration of
several $1 \times 2$, $1 \times 3$, $2 \times 1$, and $3 \times 1$
rectangular blocks arranged in an $m \times n$ grid.
(In the commercial version,
the board is $6 \times 6$, length-two rectangles are realized as \emph{cars},
and length-three rectangles are \emph{trucks}.)
Horizontally oriented blocks
can slide left and right, and vertically oriented blocks can slide up and down,
provided the blocks remain disjoint.
(Cars and trucks can drive only forward or reverse.)
The goal is to remove a particular block from the puzzle via a one-unit
opening in the bounding rectangle.
Flake and Baum \cite{Flake-Baum-2002} proved that this formulation of
Rush Hour is PSPACE-complete.
Their approach is also the basis for Nondeterministic Constraint Logic
described in Section~\ref{Constraint Logic}.
A version of Rush Hour played on a triangular grid, \emph{Triagonal Slide-Out},
is also PSPACE-complete \cite{Hearn-2006}.
Tromp and Cilibrasi \cite{Tromp-2000, Tromp-Cilibrasi-2004} strengthened
Flake and Baum's result by showing that Rush Hour remains PSPACE-complete
even when all the blocks have length two (cars).
The complexity of the problem remains open
when all blocks are $1 \times 1$ but labeled whether they move only
horizontally or only vertically
\cite{Hearn-Demaine-2002, Tromp-Cilibrasi-2004, Hearn-Demaine-2005}.
As with Subway Shuffle (Section~\ref{Moving Tokens}), solving the puzzle (by escaping the
target block from the grid) empirically seems hard \cite{Tromp-Cilibrasi-2004}, whereas
it is easy to determine whether a block may move at all by a sequence
of moves. Indeed, $1 \times 1$ Rush Hour is a restricted form of Subway Shuffle, where
there are only two colors, the graph is a grid, and horizontal edges and vertical edges
use different colors. Thus, it should be easier to find positive results for $1 \times 1$ Rush Hour,
and easier to find hardness results for Subway Shuffle. We conjecture that both are PSPACE-complete,
but existing proof techniques seem inapplicable.

\subsection{Pushing Blocks}
\label{Pushing Blocks}

Similar in spirit to the sliding-block puzzles in Section~\ref{Sliding Blocks}
are \emph{pushing-block puzzles}.  In sliding-block puzzles, an exterior agent
can move arbitrary blocks around, whereas pushing-block puzzles embed a
\emph{robot} that can only move adjacent blocks but can also move itself within
unoccupied space.  The study of this type of puzzle was initiated by Wilfong
\cite{Wilfong-1991}, who proved that deciding whether the robot can reach a
desired target is NP-hard when the robot can push and pull L-shaped blocks.

Since Wilfong's work, research has concentrated on the simpler model in which
the robot can only push blocks and the blocks are unit squares.  Types of
puzzles are further distinguished by how many blocks can be pushed at once,
whether blocks can additionally be defined to be \emph{unpushable} or
\emph{fixed} (tied to the board), how far blocks move when pushed, and the goal
(usually for the robot to reach a particular location).  Dhagat and O'Rourke
\cite{Dhagat-O'Rourke-1992}
initiated the exploration of square-block puzzles by proving that
\textsc{Push}-\texttt{*}, in which arbitrarily many blocks can be pushed at
once, is NP-hard with fixed blocks.  Bremner, O'Rourke, and Shermer
\cite{Bremner-O'Rourke-Shermer-1994} strengthened this result to
PSPACE-completeness.  Recently, Hoffmann \cite{Hoffmann-2000} proved that
\textsc{Push}-\texttt{*} is NP-hard even without fixed blocks, but it remains
open whether it is in NP or PSPACE-complete.

Several other results allow only a single block to be pushed at once.
In this context, fixed blocks are less crucial because a $2 \times 2$
cluster of blocks can never be disturbed.
A well-known computer puzzle in this context is \emph{Sokoban}, where
the goal is to place each block onto any one of the
designated target squares.  This puzzle was proved
NP-hard by Dor and Zwick \cite{Dor-Zwick-1999} and later
PSPACE-complete by Culberson
\cite{Culberson-1998}.
Later this result was strengthened to configurations with no fixed blocks
\cite{Hearn-Demaine-2002, Hearn-Demaine-2005}.
A simpler puzzle, called \textsc{Push}-$1$, arises when the goal is simply for
the robot to reach a particular position, and there are no fixed blocks.
Demaine, Demaine, and O'Rourke
\cite{Demaine-Demaine-O'Rourke-2000-push1} prove that this puzzle is
NP-hard, but it remains open whether it is in NP or PSPACE-complete.
On the other hand, PSPACE-completeness has been established for
\textsc{Push}-$2$-F, in which there are fixed blocks
and the robot can push two blocks at a time \cite{Demaine-Hearn-Hoffmann-2002}.

\begin{wrapfigure}{r}{1.5in}
  \centering
  \vspace*{-\intextsep}
  \includegraphics[scale=0.7]{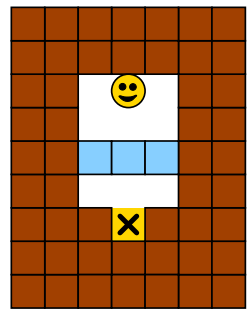}
  \caption{A \textsc{Push}-$1$ or \textsc{PushPush}-$1$ puzzle:
           move the robot to the X by pushing light blocks.}
  \label{pushing blocks}
\end{wrapfigure}

A variation on the \textsc{Push} series of puzzles, called \textsc{PushPush},
is when a block always slides as far as possible when pushed.
Such puzzles arise in a computer game with the same name
\cite{Demaine-Demaine-O'Rourke-2000-push1,Demaine-Demaine-O'Rourke-2000-pushpush2d,O'Rourke-SmithPSG-1999}.
\textsc{PushPush}-$1$ was established to be NP-hard slightly earlier than
\textsc{Push}-$1$
\cite{Demaine-Demaine-O'Rourke-2000-pushpush2d,O'Rourke-SmithPSG-1999};
the \textsc{Push}-$1$ reduction \cite{Demaine-Demaine-O'Rourke-2000-push1}
also applies to \textsc{PushPush}-$1$.
\textsc{PushPush}-$k$ was later shown PSPACE-complete for any fixed $k \geq 1$
\cite{Demaine-Hoffman-Holzer-2004}.
Hoffmann's reduction for \textsc{Push}-\texttt{*} also proves that
\textsc{PushPush}-\texttt{*} is NP-hard without fixed blocks.

Another variation, called \textsc{Push}-\texttt{X},
disallows the robot from revisiting a square (the robot's path cannot cross).
This direction was suggested in \cite{Demaine-Demaine-O'Rourke-2000-push1}
because it immediately places the puzzles in NP.
Demaine and Hoffmann \cite{Demaine-Hoffmann-2001} proved that
\textsc{Push}-$1$\texttt{X} and \textsc{PushPush}-$1$\texttt{X}
are NP-complete.  Hoffmann's reduction for \textsc{Push}-\texttt{*}
also establishes NP-completeness of \textsc{Push}-\texttt{*}\texttt{X}
without fixed blocks.

Friedman \cite{Friedman-2002} considers another variation,
where gravity acts on the blocks (but not the robot):
when a block is pushed it falls if unsupported.
He shows that \textsc{Push}-$1$-G,
where the robot may push only one block, is NP-hard.

\emph{River Crossing}, another ThinkFun puzzle
(originally \emph{Plank Puzzles} by Andrea Gilbert \cite{Gilbert-2000}),
is similar to pushing-block puzzles in that there is a unique piece
that must be used to move the other puzzle pieces.
The game board is a grid, with \emph{stumps} at some intersections, and 
\emph{planks} arranged
\begin{wrapfigure}{r}{2.5in}
  \centering
  \vspace{2ex}
  \includegraphics[width=2.5in]{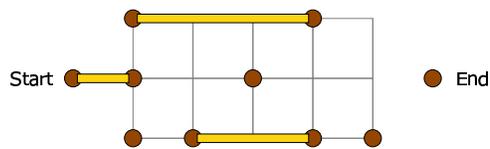}
  \caption{A River Crossing puzzle. Move from start to end.}
  \label{plank example}
\end{wrapfigure}
between some pairs of stumps, along the grid lines.
A special piece, the \emph{hiker}, always stands on some plank,
and can walk along connected planks.
He can also pick up and carry a single plank at a time,
and deposit that plank between stumps that are appropriately spaced.
The goal is for the hiker to reach a particular stump.
Figure~\ref{plank example} shows a sample puzzle.
Hearn \cite{Hearn-2004,Hearn-2006} proves that River Crossing is
PSPACE-complete, by a reduction from Constraint Logic.

\subsection{Rolling and Tipping Blocks}

In some puzzles the blocks can change their orientation as well as their
position.  Rolling-cube puzzles were popularized by Martin Gardner in his
\emph{Mathematical Games} columns in \emph{Scientific American}
\cite{g-mgc-63, g-mgc-65, g-mgc-75}.
In these puzzles, one or more cubes with some labeled sides (often dice)
are placed on a grid, and may roll from cell to cell,
pivoting on their edges between cells.
Some cells may have labels which must match the face-up label of the cube
when it visits the cell.  The tasks generally involve completing
some type of circuit while satisfying some label constraints
(e.g., by ensuring that a particular labeled face never points up).
Recently Buchin et al.~\cite{Buchin-et-al-2007} formalized this type of
problem and derived several results.
In their version, every labeled cell must be visited,
with the label on the top face of the cube matching the cell label.
Cells can be labeled, \emph{blocked}, or \emph{free}.
Blocked cells cannot be visited;
free cells can be visited regardless of cube orientation.  Such puzzles
turn out to be easy if labeled cells can be visited multiple times.
If each labeled cell must be visited exactly one,
the problem becomes NP-complete.

Rolling-block puzzles were later generalized by Richard Tucker to puzzles
where the blocks no longer need be cubes.
In these puzzles, the blocks are $k \times m \times n$ boxes.
Typically, some grid cells are blocked, and the goal is to move a block
from a start position to an end position by successive rotations into
unblocked cells.  Buchin and Buchin \cite{Buchin-Buchin-2007} recently showed 
that these puzzles are PSPACE-complete when multiple rolling blocks are used,
by a reduction from Constraint Logic.

A commercial puzzle involving blocks that tip is the ThinkFun puzzle
\emph{TipOver} (originally the \emph{Kung Fu Packing Crate Maze}
by James Stephens \cite{Stephens-2003}).  In this puzzle, all the blocks
are $1 \times 1 \times n$ (``crates'') and initially vertical.
A \emph{tipper} stands on a starting crate,
and attempts to reach a target crate.
The tipper may tip over a vertical crate it is standing on,
if there is empty space in the grid for it to fall into.
The tipper may also move between connected crates (but cannot jump diagonally).
Unlike rolling-block puzzles, in these tipping puzzles once a block has
tipped over it may not stand up again (or indeed move at all).
Hearn \cite{Hearn-2006a} showed that TipOver is NP-complete,
by a reduction from Constraint Logic.

A two-player tipping-block game inspired by TipOver, called
\emph{Cross Purposes}, was invented by Michael Albert, and named by
Richard Guy, at the Games at Dalhousie III workshop in 2004.
In Cross Purposes, all the blocks are $1 \times 1 \times 2$,
and initially vertical.  One player, \emph{horizontal}, may only tip blocks
over horizontally as viewed from above; the other player, \emph{vertical},
may only tip blocks over vertically as viewed from above.
The game follows normal play: the last player to move wins.
Hearn \cite{Hearn-2008} proved that Cross Purposes is PSPACE-complete,
by a reduction from Constraint Logic.

\begin{wrapfigure}{r}{3in}
  \centering
  \includegraphics[width=3in]{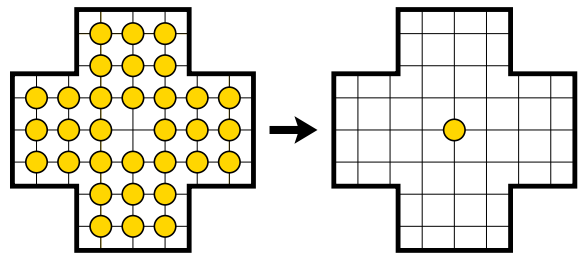}
  \caption{Central peg solitaire (\protect\hbox{Hi-Q}):
    initial and target configurations.}
  \label{peg solitaire central}
\end{wrapfigure}

\subsection{Peg Solitaire (Hi-Q)}
\label{Peg Solitaire}

The classic \emph{peg solitaire puzzle} is shown in Figure~\ref{peg solitaire
central}.
Pegs are arranged in a Greek cross, with the central peg missing.
Each move \emph{jumps} a peg over another peg (adjacent horizontally or
vertically) to the opposite unoccupied position within the cross, and removes
the peg that was jumped over.  The goal is to leave just a single peg, ideally
located in the center.
A variety of similar peg solitaire puzzles are given in \cite{Beasley-1985}.
See also Chapter 23 of \emph{Winning Ways}
\cite[pp.~803--841]{Berlekamp-Conway-Guy-2001}.

A natural generalization of peg solitaire
is to consider pegs arranged in an
$n \times n$ board and the goal is to leave a single peg.  Uehara and Iwata
\cite{Uehara-Iwata-1990} proved that it is NP-complete to decide whether such a
puzzle is solvable.

On the other hand, Moore and Eppstein \cite{Moore-Eppstein-2002} proved that
the one-dimensional special case (pegs along a line) can be solved in
polynomial time.  In particular, the binary strings
representing initial configurations that can reach a single peg
turn out to form a regular language, so they can be parsed using
regular expressions.
(This fact has been observed in various contexts;
see \cite{Moore-Eppstein-2002} for references as well as a proof.)
Using this result, Moore and Eppstein build a polynomial-time algorithm
to maximize the number of pegs removed from any given puzzle.

Moore and Eppstein \cite{Moore-Eppstein-2002} also study the natural impartial
two-player game arising from peg solitaire, \emph{duotaire}: players take
turns jumping, and the winner is determined by normal play.  (This game is
proposed, e.g., in \cite{Beasley-1985}.)  Surprisingly, the complexity of this
seemingly simple game is open.  Moore and Eppstein conjecture that the game
cannot be described even by a context-free language, and prove this conjecture
for the variation in which multiple jumps can be made in a single move.
Konane (Section~\ref{Konane}) is a natural partizan two-player game arising
from peg solitaire.

\subsection{Card Solitaire}

Two solitaire games with playing cards have been analyzed from a
complexity standpoint.  With all such games, we must generalize the
deck beyond 52 cards.  The standard approach is to keep the number of suits
fixed at four, but increase the number of ranks in each suit to~$n$.

\emph{Klondike} or \emph{Solitaire} is the classic game, in particular
bundled with Microsoft Windows since its early days.
In the perfect information of this game, we suppose the player knows
all of the normally hidden cards.
Longpr\'e and McKenzie \cite{Longpre-McKenzie-2007}
proved that the perfect-information version is NP-complete,
even with just three suits.
They also prove that Klondike with one black suit and one red suit
is NL-hard; Klondike with any fixed number of black suits and no red suits
is in NL; Klondike with one suit is in AC$^0[3]$; among other results.

\emph{FreeCell} is another common game distributed with Microsoft Windows
since XP.  We will not attempt to describe the rules here.
Helmert \cite{Helmert-2003} proved that FreeCell is NP-complete,
for any fixed positive number of free cells.

\subsection{Jigsaw, Edge-Matching, Tiling, and Packing Puzzles}
\label{Jigsaw}

\emph{Jigsaw puzzles} \cite{Williams-2004}
are another one of the most popular kinds of puzzles, dating back to the 1760s.
One way to formalize such puzzles is as a collection of square pieces,
where each side is either straight or augmented with a tab or a pocket
of a particular shape.  The goal is to arrange the given pieces so that
they form exactly a given rectangular shape.
Although this formalization does not explicitly allow for patterns on pieces
to give hints about whether pieces match, this information can simply be
encoded into the shapes of the tabs and pockets, making them compatible
only when the patterns also match.
Deciding whether such a puzzle has a solution was recently shown
NP-complete \cite{Demaine-Demaine-2007-jigsaw}.

A closely related type of puzzles is \emph{edge-matching puzzles}
\cite{Haubrich-1995}, dating back to the 1890s.
In the simplest form, the pieces are squares and, instead of tabs or pockets,
each edge is colored to indicate compatibility.
Squares can be placed side-by-side if the edge colors match,
either being exactly equal (\emph{unsigned} edge matching)
or being opposite (\emph{signed} edge matching).
Again the goal is to arrange the given pieces into a given rectangle.
Signed edge-matching puzzles are common in reality where the colors
are in fact images of lizards, insects, etc., and one side shows the
head while the other shows the tail.
Such puzzles are almost identical to jigsaw puzzles,
with tabs and pockets representing the sign; jigsaw puzzles are effectively
the special case in which the boundary must be uniformly colored.
Thus, signed edge-matching puzzles are NP-complete,
and in fact, so are unsigned edge-matching puzzles
\cite{Demaine-Demaine-2007-jigsaw}.

An older result by Berger \cite{Berger-1966} proves that the infinite
generalization of edge-matching puzzles, where the goal is to tile the entire
plane given infinitely many copies of each tile type, is undecidable.
This result is for unsigned puzzles, but by a simple reduction
in \cite{Demaine-Demaine-2007-jigsaw} it holds for signed puzzles as well.
Along the same lines,
Garey, Johnson, and Papadimitriou \cite[p.~257]{Garey-Johnson-1979}
observe that the finite version with a given target rectangle
is NP-complete when given arbitrarily many copies of each tile type.
In contrast, the finite result above requires every given tile
to be used exactly once, which corresponds more closely to real puzzles.

A related family of tiling and packing puzzles involve polyforms
such as \emph{polyominoes}, edge-to-edge joinings of unit squares.
In general, we are given a collection of such shapes and a target shape
to either tile (form exactly) or pack (form with gaps).
In both cases, pieces cannot overlap, so the tiling problem is actually
a special case in which the piece areas sum to the target areas.
One of the few positive results is for (mathematical) \emph{dominoes},
polyominoes (rectangles) made from two unit squares: the tiling and
(grid-aligned) packing problems can be solved in polynomial time
for arbitrary polyomino target shapes
by perfect and maximum matching, respectively;
see also the elegant tiling criterion of Thurston \cite{Thurston-1990}.
In contrast, with ``real'' dominoes, where each square has a color and
adjacent dominoes must match in color, tiling (and hence packing)
becomes NP-complete \cite{Biedl-2005-domino}.
The tiling problem is also NP-complete when the target shape is a polyomino
with holes and the pieces are all identical $2 \times 2$ squares,
or $1 \times 3$ rectangles, or $2 \times 2$ L shapes \cite{Moore-Robson-2001}.
The packing problem \cite{Li-Cheng-1989} and the tiling problem
\cite{Demaine-Demaine-2007-jigsaw} are NP-complete
when the given pieces are differently sized squares
and the target shape is a square.
Finally, the tiling problem is NP-complete when the
given pieces are polylogarithmic-area polyominoes
and the target shape is a square \cite{Demaine-Demaine-2007-jigsaw};
this result follows by simulating jigsaw puzzles.

\subsection{Minesweeper}

\emph{Minesweeper} is a well-known imperfect-information computer puzzle
popularized by its inclusion in Microsoft Windows.  Gameplay takes place on an
$n \times n$ board, and the player does not know which squares contain mines.
A move consists of uncovering a square; if that square contains a mine, the
player loses, and otherwise the player is revealed the number of mines in the
$8$ adjacent squares.  The player also knows the total number of mines.

There are several problems of interest in Minesweeper.  For example, given a
configuration of partially uncovered squares (each marked with the number of
adjacent mines), is there a position that can be safely uncovered?  More
generally, what is the probability that a given square contains a mine,
assuming a uniform distribution of remaining mines?  A different generalization
of the first question is whether a given configuration is \emph{consistent},
i.e., can be realized by a collection of mines.  A consistency checker would
allow testing whether a square can be guaranteed to be free of mines, thus
answering the first question.  An additional problem is to decide whether a given
configuration has a unique realization.

Kaye \cite{Kaye-2000} proves that testing consistency is NP-complete.
This result leaves open the complexity of the other questions mentioned above.
Fix and McPhail \cite{Fix-McPhail-2004} strengthen Kaye's result to show
NP-completeness of determining consistency when the uncovered numbers
are all at most~$1$.
McPhail \cite{McPhail-2003} also shows that, given a consistent placement of
mines, determining whether there is another consistent placement is
NP-complete (ASP-completeness from Section~\ref{Pencil-and-Paper}).

Kaye \cite{Kaye-2000-infinite} also proves that an infinite generalization of
Minesweeper is undecidable.  Specifically, the question is whether a given
finite configuration can be extended to the entire plane.  The rules permit a
much more powerful level of information revealed by uncovering squares; for
example, discovering that one square has a particular label might imply that
there are exactly $3$ adjacent squares with another particular label.  (The
notion of a mine is lost.)
The reduction is from tiling (Section~\ref{Jigsaw}).

Hearn \cite{Hearn-2006, Hearn-2008a} argues that the ``natural'' decision question
for Minesweeper, in keeping with the standard form for other puzzle complexity results,
is whether a given (assumed consistent) instance can (definitely) be solved, which
is a different question from any of the above. He observes that a simple modification
to Kaye's construction shows that this question is coNP-complete, an unusual
complexity class for a puzzle. The reduction is from Tautology. (If the instance is not known
to be consistent, then the problem may not be in coNP.) Note that this question is not 
the same as whether a given configuration has a unique realization: there could be
multiple realizations, as long as the player is guaranteed that known-safe moves will
eventually reveal the entire configuration.

\subsection{Mahjong Solitaire (Shanghai)}

\emph{Majong solitaire} or \emph{Shanghai} is a common computer game played
with Mahjong tiles, stacked in a pattern that hides some tiles,
and shows other tiles, some of which are completely exposed.
Each move removes a pair of matching tiles that are completely exposed;
there are precisely four tiles in each equivalence class of matching.
The goal is to remove all tiles.

Condon, Feigenbaum, Lund, and Shor \cite{Condon-Feigenbaum-Lund-Shor-1997}
proved that it is PSPACE-hard to approximate the maximum probability of
removing all tiles within a factor of $n^\epsilon$, assuming that there are
arbitrarily many quadruples of matching tiles and that the hidden tiles are
uniformly distributed.  Eppstein \cite{Eppstein-cgt-hard} proved that
it is NP-complete to decide whether all tiles can be removed in the
perfect-information version of this puzzle where all tile positions are known.

\subsection{Tetris}

\emph{Tetris} is a popular computer puzzle game invented in the mid-1980s by
Alexey Pazhitnov, and by 1988 it became the best-selling game in the
United States and England.
The game takes place in a rectangular grid (originally, $20 \times 10$)
with some squares occupied by blocks.
During each move, the computer generates a tetromino piece stochastically
and places it at the top of the grid; the player can rotate the piece and
slide it left or right as it falls downward.  When the piece hits another
piece or the floor, its location freezes and the move ends.
Also, if there are any completely filled rows, they disappear,
bringing any rows above down one level.

To make Tetris a perfect-information puzzle,
Breukelaar et al.~\cite{Breukelaar-Demaine-Hohenberger-Hoogeboom-Kosters-Liben-Nowell-2004}
suppose that the player knows in advance the entire sequence of pieces
to be delivered.  Such puzzles appear in \emph{Games Magazine}, for example.
They then prove NP-completeness of deciding whether it is possible to
stay alive, i.e., always be able to place pieces.
Furthermore, they show that maximizing various notions of score,
such as the number of lines cleared, is NP-complete to approximate
within an $n^{1-\varepsilon}$ factor.
The complexity of Tetris remains open with a constant number of rows or
columns, or with a stochastically chosen piece sequence as in
\cite{Papadimitriou-1985}.

\subsection{Clickomania (Same Game)}

\begin{figure}
  \centering
  \includegraphics[width=0.675\linewidth]{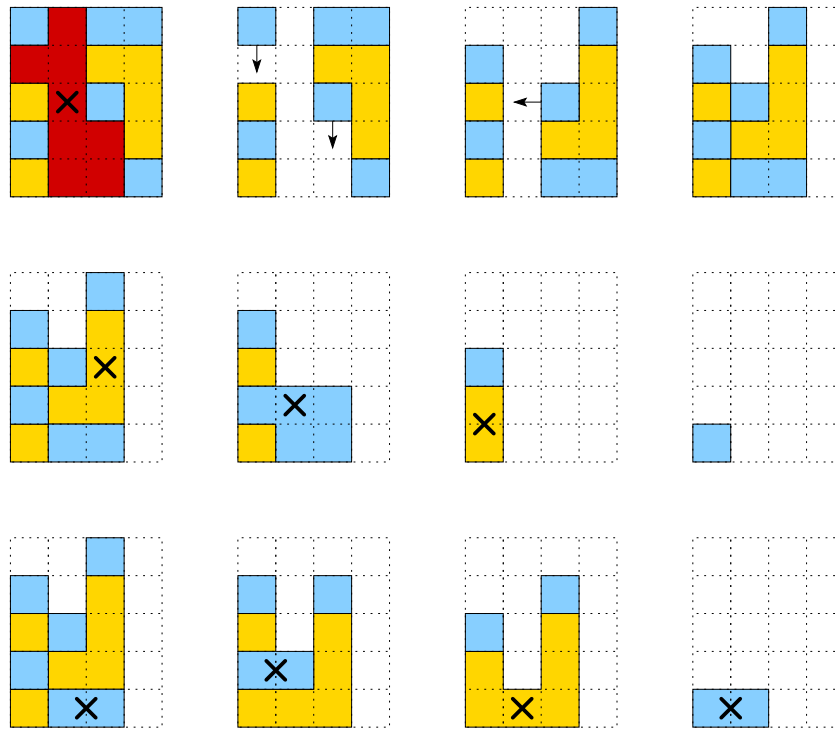}
  \caption{The falling rules for removing a group in Clickomania (top),
    a failed attempt (middle), and a successful solution (bottom).}
  \label{clickomania example}
\end{figure}

\emph{Clickomania} or \emph{Same Game}
\cite{Biedl-Demaine-Demaine-Fleischer-Jacobsen-Munro-2002} is a computer puzzle
consisting of a rectangular grid of square blocks each colored one of $k$
colors.  Horizontally and vertically adjacent blocks of the same color are
considered part of the same \emph{group}.  A move selects a group containing at
least two blocks and removes those blocks, followed by two ``falling'' rules;
see Figure~\ref{clickomania example} (top).  First, any blocks remaining
above created holes fall down in each column.  Second, any empty columns are
removed by sliding the succeeding columns left.

The main goal in Clickomania is to remove all the blocks.
A simple example for
which this is impossible is a checkerboard, where no move can be made.  A
secondary goal is to maximize the score, typically defined by $k^2$ points
being awarded for removal of a group of $k$ blocks.

Biedl et al.\ \cite{Biedl-Demaine-Demaine-Fleischer-Jacobsen-Munro-2002} proved
that it is NP-complete to decide whether all blocks can be removed in a
Clickomania puzzle.
This complexity result holds even for puzzles with two
columns and five colors, and for puzzles with five columns and three colors.
On the other hand, for puzzles with one column (or, equivalently, one row) and
arbitrarily many colors, they show that the maximum number of blocks can be
removed in polynomial time.
In particular, the puzzles whose blocks can all be
removed are given by the context-free grammar $S \to \Lambda\,|\,SS\,|\,c S
c\,|\,cScSc$ where $c$ ranges over all colors.

Various cases of Clickomania remain open, for example, puzzles with two colors,
and puzzles with $O(1)$ rows.  Richard Nowakowski suggested a two-player
version of Clickomania, described in
\cite{Biedl-Demaine-Demaine-Fleischer-Jacobsen-Munro-2002}, in which players
take turns removing groups and normal play determines the winner; the
complexity of this game remains open.

A related puzzle is called \emph{Vexed}, also \emph{Cubic}. In this puzzle there
are fixed blocks, as well as the mutually annihilating colored blocks. A move in
Vexed is to slide a colored block one unit left or right into an empty space, whereupon
gravity will pull the block down until it contacts another block; then any touching blocks
of the same color disappear. Again the goal is to remove all the colored blocks.
Friedman \cite{Friedman-2001} showed that Vexed is NP-complete.\footnote{
David Eppstein pointed out that all that was shown was NP-hardness; the problem was not
obviously in NP (\url{http://www.ics.uci.edu/~eppstein/cgt/hard.html}). Friedman and R. Hearn
together showed that it is in NP as well (personal communication).}

\subsection{Moving Coins}

\begin{figure}
  \def\coinscale{0.8}%
  \centering
  \begin{tabular}{ccc}
  \subfigure[Turn the pyramid upside-down in three moves.]
    {\includegraphics[scale=\coinscale]{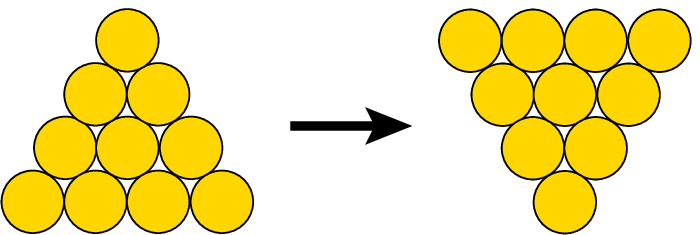}} & \hfil\hfil &
  \subfigure[Re-arrange the pyramid into a line in seven moves.]
    {\includegraphics[scale=\coinscale]{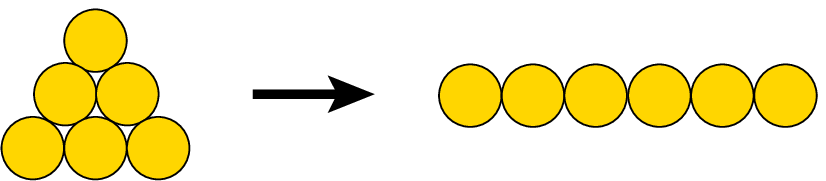}} \\
  \subfigure[Flip the diagonal in $18$ moves.]
    {\includegraphics[scale=\coinscale]{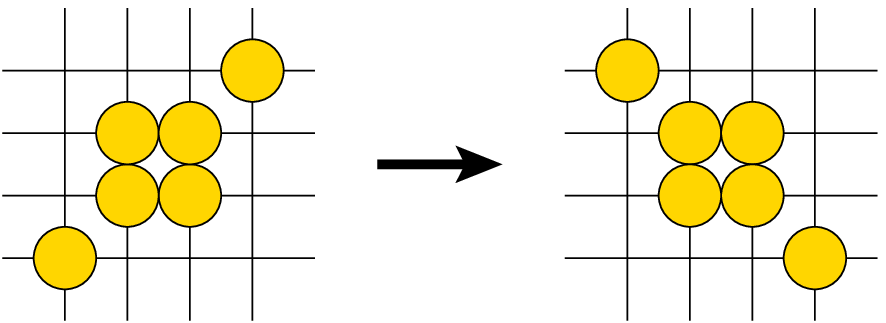}} & \hfil\hfil &
  \subfigure[Invert the V in $24$ moves.]
    {\includegraphics[scale=0.65]{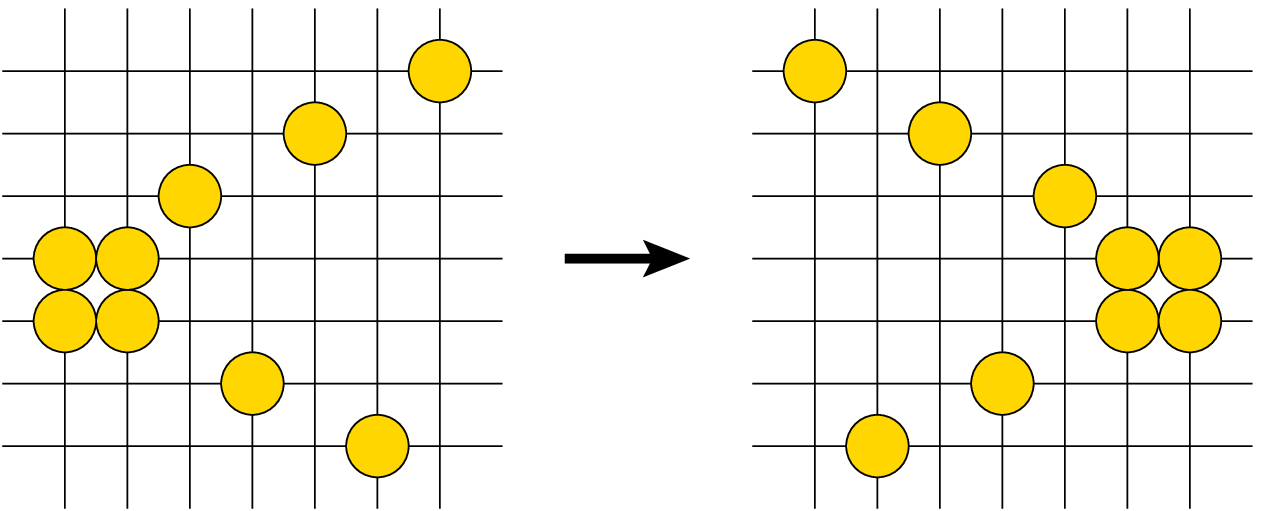}}
  \end{tabular}%
  \caption{Coin-moving puzzles in which each move places a coin adjacent to two
    other coins; in the bottom two puzzles, the coins must also remain on the
    square lattice.  The top two puzzles are classic, whereas the bottom two
    puzzles were designed in \protect\cite{Demaine-Demaine-Verrill-2000}.}
  \label{moving coin examples}
\end{figure}

Several coin-sliding and coin-moving puzzles fall into the following general
framework: re-arrange one configuration of unit disks in the plane into another
configuration by a sequence of moves, each repositioning a coin in an empty
position that touches at least two other coins.  Examples of such puzzles
are shown in Figure~\ref{moving coin examples}.  This framework can be
further generalized to nongeometric puzzles involving movement of tokens on
graphs with adjacency restrictions.

Coin-moving puzzles are analyzed by Demaine, Demaine, and Verrill
\cite{Demaine-Demaine-Verrill-2000}.  In particular, they study puzzles as in
Figure~\ref{moving coin examples} in which the coins' centers remain on either
the triangular lattice or the square lattice.  Surprisingly, their results for
deciding solvability of puzzles are positive.

For the triangular lattice, nearly all puzzles are solvable, and there is a
polynomial-time algorithm characterizing them.  For the square lattice, there
are more stringent constraints.  For example, the bounding box cannot increase
by moves; more generally, the set of positions reachable by moves given an
infinite supply of extra coins (the \emph{span}) cannot increase.  Demaine,
Demaine, and Verrill show that, subject to this constraint, there is a
polynomial-time algorithm to solve all puzzles with at least two extra coins
past what is required to achieve the span.  (In particular, all such puzzles
are solvable.)

\subsection{Dyson Telescopes}

The \emph{Dyson Telescope Game} is an online puzzle produced by the 
Dyson corporation, whimsically based on their telescoping vacuum cleaners.
The goal is to maneuver a ball on a square grid from a starting position to a
goal position by extending and retracting telescopes on the grid.
When a telescope is extended, it grows to its maximum length in the direction
it points (parameters of each telescope), unless it is stopped by another
telescope.
If the ball is in the way, it is pushed by the end of the telescope.
When a telescope is retracted, it shrinks back to unit length,
pulling the ball with it if the ball was at the end of the telescope.

Demaine et al.~\cite{Demaine-et-al-2008} showed that determining whether a 
given puzzle has a solution is PSPACE-complete in the general case.
On the other hand, the problem is polynomial for certain restricted
configurations which are nonetheless interesting for humans to play.
Specifically, if no two telescopes face each other and overlap when extended
by more than one space, then the problem is polynomial.
Many of the game levels in the online version have this property.

\subsection{Reflection Puzzles}

Two puzzles involving reflection of directional light or motion
have been studied from a complexity-theoretic standpoint.

In \emph{Reflections} \cite{Kempe-2003}, we are given a rectangular grid
with one square marked with a laser pointed in one of the four axis-parallel
directions, one or more squares marked as light bulbs, some squares
marked one-way in an axis-parallel direction, and remaining squares
marked either empty or wall.
We are also given a number of diagonal mirrors and/or T-splitters
which we can place arbitrarily into empty squares.
The light then travels from the laser; when it meets a diagonal mirror,
it reflects by $90^\circ$ according to the orientation of the mirror;
when it meets a splitter at the base of the T, it splits into both
orthogonal directions; when it meets a one-way square, it stops unless
the light direction matches the one-way orientation;
when it meets a light bulb, it toggles the bulb's state and stops;
and when it meets a wall, it stops.
The goal is to place the mirrors and splitters so that each light bulb
gets hit an odd number of times.
This puzzle is NP-complete \cite{Kempe-2003}.

In \emph{Reflexion} \cite{Holzer-Schwoon-2004-Reflexion},
we are given a rectangular grid in which squares are either walls,
mirrors, or diamonds.  Also, one square is the starting position
for a ball and another square is the target position.
We may release the ball in one of the four axis-parallel directions,
and we may flip mirrors between their two diagonal orientations
while the ball moves.  The ball travels like a ray of light,
reflecting at mirrors and stopping at walls; at diamonds, it turns around
and erases the diamond.  The goal is to reach the target position.
In this simplest form, Reflexion is SL-complete which actually implies
a polynomial-time algorithm \cite{Holzer-Schwoon-2004}.
If some of the mirrors can be flipped only before the ball releases,
the puzzle becomes NP-complete.
If some trigger squares toggle other squares between wall and empty,
or if some squares contain horizontally or vertically movable blocks
(which also cause the ball to turn around), then the puzzle becomes
PSPACE-complete.

\subsection{Lemmings}

\emph{Lemmings} is a popular computer puzzle game dating back to the early
1990s.  Characters called lemmings start at one or more initial locations
and behave deterministically according to their mode, initially
just walking in a fixed direction, turning around at walls,
and falling off cliffs, dying if it falls too far.
The player can modify this basic behavior by applying a skill to a lemming;
each skill has a limited number of such applications.
The goal is for a specified number of lemmings to reach a specified
target position.  The exact rules, particularly the various skills,
are too complicated to detail here.
Cormode \cite{Cormode-2004-lemmings} proved that such puzzles are
NP-complete, even with just one lemming.
Membership in NP follows from assuming a polynomial upper bound on
the time limit in a level (a fairly accurate modeling of the actual game);
Cormode conjectures that this assumption does not affect the result.


\section{Cellular Automata and Life}
\label{Cellular Automata and Life}

Conway's \emph{Game of Life} is a zero-player cellular automaton played on the
square tiling of the plane.  Initially, certain cells (squares) are marked
\emph{alive} or \emph{dead}.  Each move globally evolves the cells: a live cell
remains alive if between $2$ and $3$ of its $8$ neighbors were alive,
and a dead cell becomes alive if it had precisely $3$ live neighbors.

Many questions can be asked about an initial configuration of Life; one key
question is whether the population will ever completely die out (no cells are
alive).  Chapter 25 of \emph{Winning Ways}
\cite[pp.~927--961]{Berlekamp-Conway-Guy-2001} describes a reduction showing
that this question is undecidable.  In particular, the same question
about Life restricted within a polynomially bounded region is PSPACE-complete.
More recently, Rendell \cite{Rendell-2005} constructed an explicit
Turing machine in Life, which establishes the same results.

There are other open complexity-theoretic questions about Life.%
\footnote{These two questions were suggested by David Eppstein.}
How hard is it to tell whether a configuration is a Garden of Eden,
that is, cannot be the state that results from another?
Given a rectangular pattern in Life, how hard is it to extend the
pattern outside the rectangle to form a Still Life
(which never changes)?

Several other cellular automata, with different survival and birth rules,
have been studied; see, e.g., \cite{Wolfram-1994}.

%


\section{Open Problems}
\label{Open Problems}

Many open problems remain in Combinatorial Game Theory.
Guy and Nowakowski \cite{Guy-Nowakowski-2002}
have compiled a list of such problems.

Many open problems also remain on the algorithmic side,
and have been mentioned throughout this paper.
Examples of games and puzzles whose complexities remain unstudied,
to our knowledge, are
Domineering (Section~\ref{Domineering}),
Connect Four,
Pentominoes,
Fanorona,
Nine Men's Morris,
Chinese checkers,
Lines of Action,
Chinese Chess,
Quoridor, and
Arimaa.
For many other games and puzzles, such as Dots and Boxes
(Section~\ref{Dots and Boxes}) and pushing-block puzzles
(Section~\ref{Pushing Blocks}),
some hardness results are known, but the exact complexity remains unresolved.
It would also be interesting to consider games of imperfect information
that people play, such as Scrabble
(Section~\ref{Crossword Puzzles and Scrabble},
Backgammon, and Bridge.
Another interesting direction for future research is to build a
more comprehensive theory for analyzing combinatorial puzzles.


\section*{Acknowledgments}

Comments from several people have helped make this survey more
comprehensive, including
Martin Demaine,
Azriel Fraenkel,
Martin Kutz,
and
Ryuhei Uehara. 


\let\realbibitem=\bibitem
\def\bibitem{\par \vspace{-1.2ex}\realbibitem}

\bibliography{auctions,cellularautomata,coinsliding,combinatorialgames,complexity,jigsaw,motionplanning,internet,parameterizedcomplexity,pushingblocks,surreals,tiling,morerefs}
\bibliographystyle{alpha}

\end{document}